\documentclass[aps, pra,twocolumn,showpacs,amsmath,amssymb,superscriptaddress, longbibliography, 10pt]{revtex4-1}

\usepackage{tgtermes}
\usepackage[utf8]{inputenc}
\usepackage{bm}
\usepackage{color}
\usepackage{makeidx}
\usepackage{amsfonts}
\usepackage{amssymb}
\usepackage{amsthm}
\usepackage{graphicx}
\usepackage{epstopdf}
\usepackage{subfigure}
\usepackage{ulem}
\usepackage{amsmath,amssymb}
\usepackage{mathrsfs}
\usepackage[colorlinks=true,citecolor=blue,urlcolor=blue]{hyperref}
\usepackage{color}

\newcommand{\colo}[1]{\textcolor{blue}{#1}}

\newcommand{\bea}{\begin{eqnarray}}
\newcommand{\ea}{\end{eqnarray}}
\newcommand{\eea}{\end{eqnarray}}

\newcommand{\sumint}[1]

\begin{document}

\title{Quench dynamics of Bose-Einstein condensates in boxlike traps}
\author{Rong Du}
\affiliation{Shaanxi Key Laboratory for Theoretical Physics Frontiers, Institute of Modern Physics, Northwest University, Xi'an, 710127, China}
\author{Jian-Chong Xing}
\affiliation{Shaanxi Key Laboratory for Theoretical Physics Frontiers, Institute of Modern Physics, Northwest University, Xi'an, 710127, China}
\author{Bo Xiong}
\affiliation{School of Science, Wuhan University of Technology, Wuhan 430070, China}
\author{Jun-Hui Zheng}
\affiliation{Shaanxi Key Laboratory for Theoretical Physics Frontiers, Institute of Modern Physics, Northwest University, Xi'an, 710127, China}
\author{Tao Yang}
\email{yangt@nwu.edu.cn}
\affiliation{Shaanxi Key Laboratory for Theoretical Physics Frontiers, Institute of Modern Physics, Northwest University, Xi'an, 710127, China}
\affiliation{NSFC-SPTP Peng Huanwu Center for Fundamental Theory, Xi'an 710127, China}

\date{\today}

\begin{abstract}
  We investigate the nonequilibrium dynamics of two-dimensional Bose-Einstein condensates in boxlike traps with power-law potential boundaries by quenching the interatomic interactions. For both concave and convex potentials, we show that ring dark solitons can be excited during the quench dynamics. The modulation strength of the quench and the steepness of the boundary are two main factors affecting the evolution of the system. Five dynamic regimes are identified concerning the number of ring dark solitons excited in the condensate. For the situation without ring dark soliton excitations, the condensate undergoes damped radius oscillation. As far as the appearance of ring dark solitons, interesting structures arise from their decay. For the concave potential, the excitation patterns show a nested structure of vortex-antivortex pairs. For the convex potential, on the other hand, the dynamic excitation patterns display richer structures that have multiple transport behaviors.
\end{abstract}


\maketitle

Nonequilibrium dynamics is an active area of research ubiquitous across wide areas ranging from the early universe to condensed matter physics. New quantum physics such as Mott transitions \cite{Rev.Mod.Phys.89.011004}, synthetic gauge fields \cite{Rev.Mod.Phys.83.1523}, and topological states \cite{Nature.515.237,CPL.37.060301} have been discovered in driven systems far away from equilibrium. Bose-Einstein condensate (BEC), owing to its high tunability in experiments, provides an excellent platform to study nonequilibrium quantum dynamics. Quantum quenches, the temporal evolution following a sudden or slow change of the coupling constants of the system Hamiltonian, have attracted particular interests \cite{Rev.Mod.Phys.83.863}. A sudden change of the scattering length has been achieved experimentally recently \cite{Nat.Phys.10.116}. In atomic BECs, many fascinating nonequilibrium dynamics are involved with tuning interatomic interaction, such as the formation of matter-wave solitons \cite{science.1071021,Nguyen422,PhysRevA.70.033607,CPL.25.2370}, the transformation between different types of vector solitons \cite{PhysRevA.79.013423}, controlled collapse \cite{PhysRevA.67.013605}, Faraday patterns \cite{PhysRevLett.89.210406}, collective excitation modes \cite{PhysRevA.81.053627,PhysRevA.84.013618} and matter-wave jet emission \cite{Nature.551.356,PhysRevLett.121.243001, PhysRevA.99.063624, PhysRevA.101.031601}.

Solitons and vortices provide an avenue for exploring manifestly nonlinear properties of interacting Bose gases. These important topological excitations are closely related to nonlinear optics \cite{Phys.Rep.298.81,Opt.Commun.152.198}, superfluidity \cite{Science.290.777,PhysRevA.102.063318}, and superconductivity \cite{Kopnin_2002}. They can be created in trapped BECs by phase imprinting via carefully controlled laser fields \cite{PRL.83.5198,Science.287.97,PhysRevA.101.053629}, controlling the condensate density via creating shock waves \cite{Science.293.663}, or colliding separated condensates \cite{PhysRevLett.101.130401,PhysRevA.87.023603,PhysRevA.88.043602}. Recently, quench of the isotropic $s$-wave scattering length was employed to prepare solitons in quasi-one-dimensional box-trapped BECs \cite{PhysRevResearch.2.043256}. {The stability of dark solitons under nonresonant {$\mathcal{P}\mathcal{T}$}-symmetric pumping was studied in Ref.\,\cite{CPL.37.040502}. The motion of solitons in one-dimensional BECs under spin-orbit coupling has been proved to be governed by a nonlinear Bloch equation \cite{PhysRevA.94.061602}.}
Dark solitons and vortices are related in higher dimensions where dark solitons are prone to break up due to transverse (snaking) instability \cite{Phys.Rep.331.117}, and leading to spontaneous emergence of vortices. Ring dark soliton (RDS) is a special structure in two-dimensional (2D) systems of interest in its own right. The decay of single ring dark soliton (S-RDS) initially sitting in the condensate has been studied \cite{PRL.90.120403,PhysRevA.92.033611, PhysRevA.85.063617}. {The dynamics and modulation of RDSs in 2D BECs with tunable interaction have been studied in Ref. \cite{PhysRevA.79.023619}.} The interplay of double RDS (D-RDS) initially seeded in a condensate cloud were found to be unstable and led to the formation of complex multivortex-lattice configurations \cite{PhysRevA.104.023314}. A sudden quench of the harmonic well into a wider cylindrical well was proposed to excite RDSs in trapped 2D BECs \cite{PhysRevA.76.063606}.

In addition, the trapping geometries is another important factor affecting the dynamics of the trapped BECs. With the development of experimental technologies, traps with either hard or soft boundaries can be achieved, such as harmonic traps, hard wall box potentials \cite{PhysRevLett.110.200406}, and soft wall potential with power-law boundaries \cite{PhysRevA.82.023613}. A matter-wave dark soliton incident upon a hard-wall potential can emit a significant fraction of its energy \cite{Parker_2003}. The quantum reflection of matter-wave bright solitons from a soft wall of the form a positive $\tanh$ potential is very sensitive to the width of the potential \cite{Cornish2009}. The nonlinear Talbot effect of BEC can occur in a one-dimensional hard-wall box potential \cite{PhysRevA.63.043613}. Numerical studies have demonstrated that the formation of large-scale Onsager vortex clusters is prevalent in steep-walled traps, but suppressed in harmonic traps \cite{PhysRevLett.113.165302, PhysRevA.93.043614}. {There are many studies about 2D dark solitons in BECs with different traps \cite{PRA.67.023604,PRE.74.066615,LP.29.015501,FOP.6.46,CPB.30.120303,PRA.95.043618,PRA.99.053619}.}

In this letter, we investigate the effect of a sudden quench of the interatomic interaction on the nonequilibrium dynamics of a 2D BEC confined in different kinds of power-law traps. We identify five dynamic regimes according to the ability of exciting RDSs in the system and depict the phase diagram with respect to the modulation strength and steepness of the boundary. For relative weak modulations of interatomic interaction, the system undergoes low-lying breathing mode \cite{JPB.47.035302} with damped oscillation of the radius of the condensate. While for a stronger tuning, RDSs are excited and develop into different patterns of vortex pairs, depending on the convexity and concavity of the boundary of the trap.

\begin{figure}[t]
\begin{center}
\includegraphics[angle=0,width=0.45\textwidth]{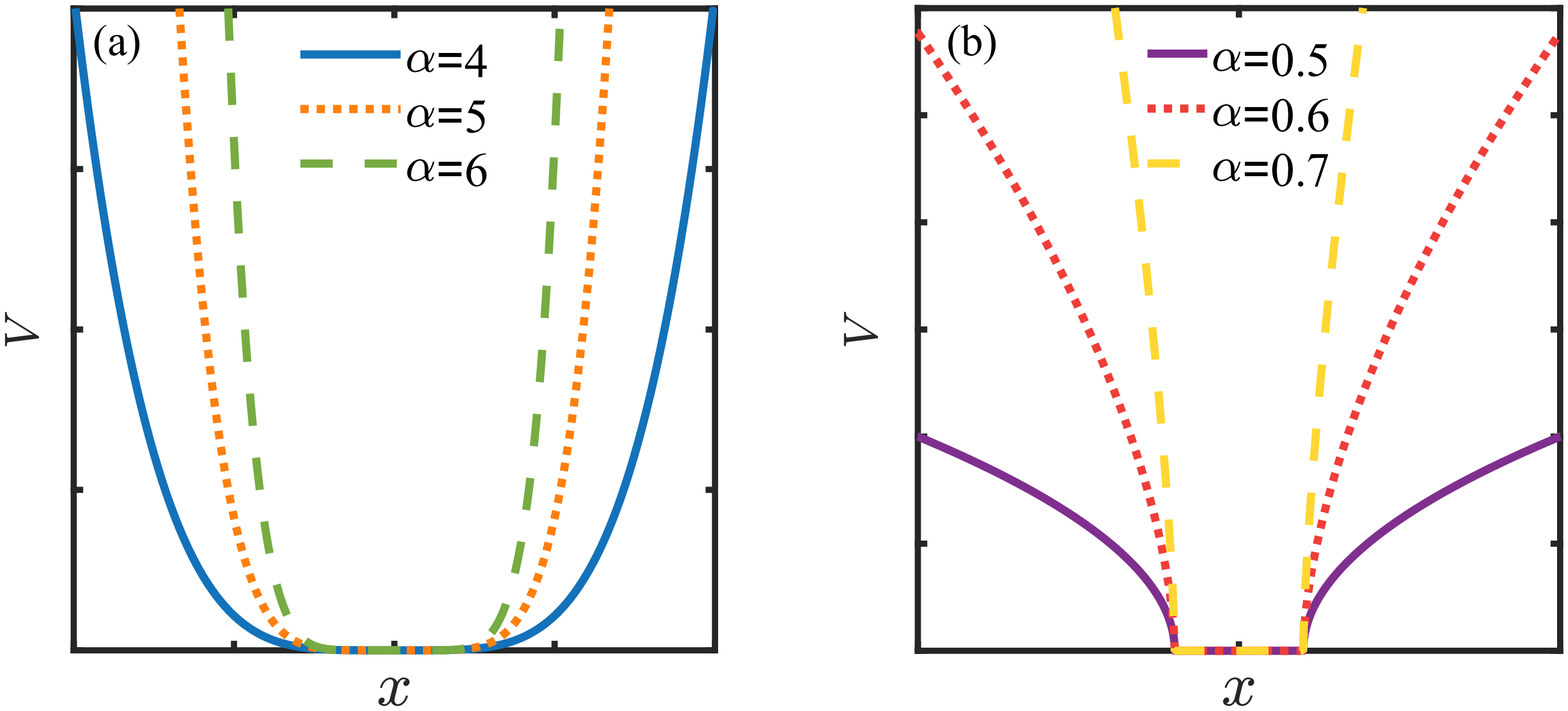}
\caption{(Color online) Schematic side-view ($y=0$) of power-law traps with different exponents $\alpha$. (a) concave potentials with $\alpha>1$. (b) convex potentials with $0< \alpha<1$. {Here $w=0.4$ and $b=2$ for (a) and $w=1.2$ and $b=0.0002$ for (b).}}
\label{Fig1}
\end{center}
\end{figure}
	
To explore the boundary effects of the trap, we consider a condensate confined in a cylindrically symmetric boxlike trap. The boundary of the trap increases as a power of the spatial coordinate of the form \cite{PhysRevA.95.013628},
\begin{eqnarray}\label{Eq1}
V(\bf{r})=\left\{ \begin{array}{cc}~~~0~~~~~~~~~~~~~~r\leq w\\
\left(\frac{r-{w}}{b}\right)^\alpha~~~~~~ r>w~ \end{array}, \right.
\end{eqnarray}
where $r=\sqrt{x^2+y^2}$. Such a trap can be achieved experimentally by two crossing blue-detuned Laguerre-Gaussian optical beams \cite{1998High}. The width of the flat part of the potential and the relative width of the potential boundary can be adjusted by ${w}$ and $b$, respectively. The exponent $\alpha$ denotes the steepness of the boundary, i.e., the sharp spatial variation of the edge potential of the boxlike trap. For $\alpha>1$, the boundary of the potential is concave, while for $0<\alpha<1$ the boundary of the potential is convex as shown in Fig.\,\ref{Fig1}. The trap tends to be an infinite square well with width $2(w+b)$ as $\alpha \rightarrow \infty$, and a finite square well with width $2w$ when $\alpha \rightarrow 0$.

The dynamics of a 2D BEC containing $N$ atoms are described by the Gross-Pitaevskii (GP) equation,
\begin{equation}\label{Eq2}
i\hbar\frac{\partial{\psi}}{\partial{t}}=-\frac{\hbar^2}{2m}\nabla^2\psi+V({\bf r})\psi+ c \colo{(t)} g N |\psi|^2\psi,
\end{equation}
where $g=2\sqrt{2\pi}\hbar \omega_z a_s a_z$ is the effective interatomic interaction strength with $a_s$ being the $s$-wave scattering length and $a_z=\sqrt{\hbar/m\omega_z}$ being the characteristic length in $z$ direction. We choose $a_s=5.4~\rm{nm}$ and $m=1.44\times10^{-25}~\rm{kg}$, appropriate to a $^{87}$Rb condensate. The trap frequency $\omega_z$ in the $z$ direction is set to be $2\pi\times100~\rm{Hz}$ to make sure that $\hbar\omega_z$ is much larger than the chemical potential of the system. The initial state of the system is the ground state of the condensate with the modulation factor $c=1$ trapped in the given box-like potential with soft-wall boundaries, which can be obtained by the imaginary time evolution. In the real time evolution, the nonequilibrium dynamics is triggered by a instantaneous quench of the interatomic interaction by changing $c$, which can be achieved experimentally via Feshbach resonances \cite{Rev.Mod.Phys.82.1225}. In the following calculation, we set $\omega_0=2\pi\times5~\rm{Hz}$ and use $a_0=\sqrt{\hbar/m\omega_0}$, $t_0=1/\omega_0$, and $E_0=\hbar\omega_0$ as units of length, time, and energy, respectively. All physics quantities in this letter are presented in dimensionless form.

\begin{figure}[tbp]
\begin{center}
\includegraphics[angle=0,width=0.48\textwidth]{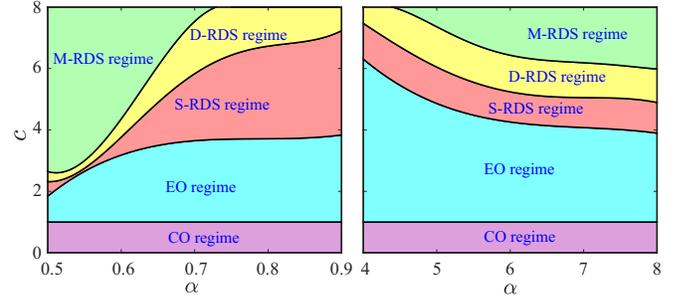}
\caption{(Color online) Phase diagram describing the dynamic regimes of the ground-state BEC suddenly quenched from ($\alpha$, $g$) to ($\alpha$, $c g$). Five dynamic regimes, i.e.,  compressing oscillating (CO), expanding oscillating (EO), single ring dark soliton (S-RDS), double ring dark solitons (D-RDS), and multiple ring dark solitons (M-RDS) are indicated in different colors. }
\label{Fig2}
\end{center}
\end{figure}

\begin{figure}[b]
\begin{center}
\includegraphics[angle=0,width=0.5\textwidth]{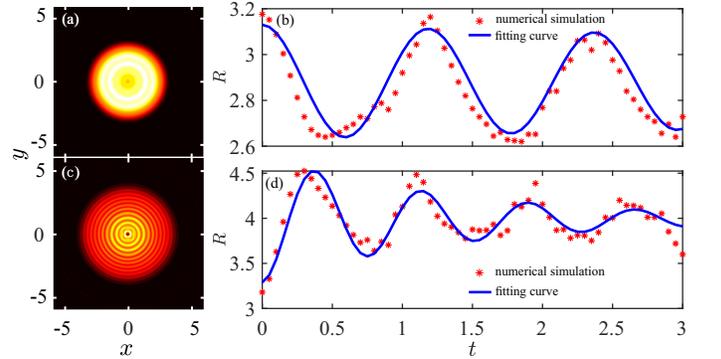}
\caption{(Color online) Hot-scale plots of atom density (white = high) and radius oscillation of the condensate. (a)-(b) $\alpha=4$ and $c=0.4$. (c)-(d) $\alpha=4$ and $c=5$ . Here $w=0.4$ and $b=2$.}
\label{Fig3}
\end{center}
\end{figure}

For both concave and convex potentials, we identify five dynamical regimes for each given $\alpha$ according to the ability of generating RDSs, which are shown in the dynamic phase diagram (Fig.\,\ref{Fig2}). We can see clearly that the phase boundaries depend strongly on the steepness of the boundary of the trap. We note that the boundary ($c=1$) is the reference line, which indicates the initial interaction strength before quench.
When we tune the modulation factor $c$ to be $0<c<1$, the reduced interatomic interaction causes the condensate cloud to compress to a minimal radius and then expand. The oscillation of the radius of the condensate is the breathing mode excited by the quench. During this dynamical process, the ring-shaped self-interference patterns are excited after the condensate is reflected by the boundary. We label this regime as compressing oscillating (CO) regime. In Fig.\,\ref{Fig3}\,(a), we show the typical density profile of the condensate in the CO regime at $t=1.25$. The radius oscillation of the condensate is shown in Fig.\,\ref{Fig3}\,(b), where the red stars are numerical data and the blue solid line is the fitting curve  $R(t) = 2.88 + 0.25 \exp(-0.06 t)\cos(5.3t)$. For $c>1$, the quench dynamics can also display a breathing mode. However, in this case, the size oscillation of the condensate starts with increasing radius, which is labeled as expanding oscillating (EO) regime. The induced density oscillations in the EO regime are much stronger than those in the CO regime as shown in Fig.\,\ref{Fig3}\,(c), due to the stronger interatomic interaction in the real time evolution. The radius oscillation of the condensate is fitted by the blue solid line $R(t) = 3.9892 - 0.7 \exp(-0.7 t)\cos(8.243t)$ in Fig.\,\ref{Fig3}\,(d). Both the damping rate and the oscillation frequency are larger than that in the CO regime, and quickly reach a steady state. In these two regimes, there are no topological excitations in the system. With increasing $c$, the quench process triggers the formation of S-RDS, D-RDS, and multiple ring dark solitons (M-RDS). The boundaries of the dynamic phases show opposite monotonic behavior on $\alpha$, depending on the convexity and concavity of the trap potential as shown in Fig.\,\ref{Fig2}. Moreover, the subsequently decay of the ring dark soliton(s) exhibites distinct mechanisms.

\begin{figure}[htp]
\begin{center}
\includegraphics[angle=0,width=0.42\textwidth]{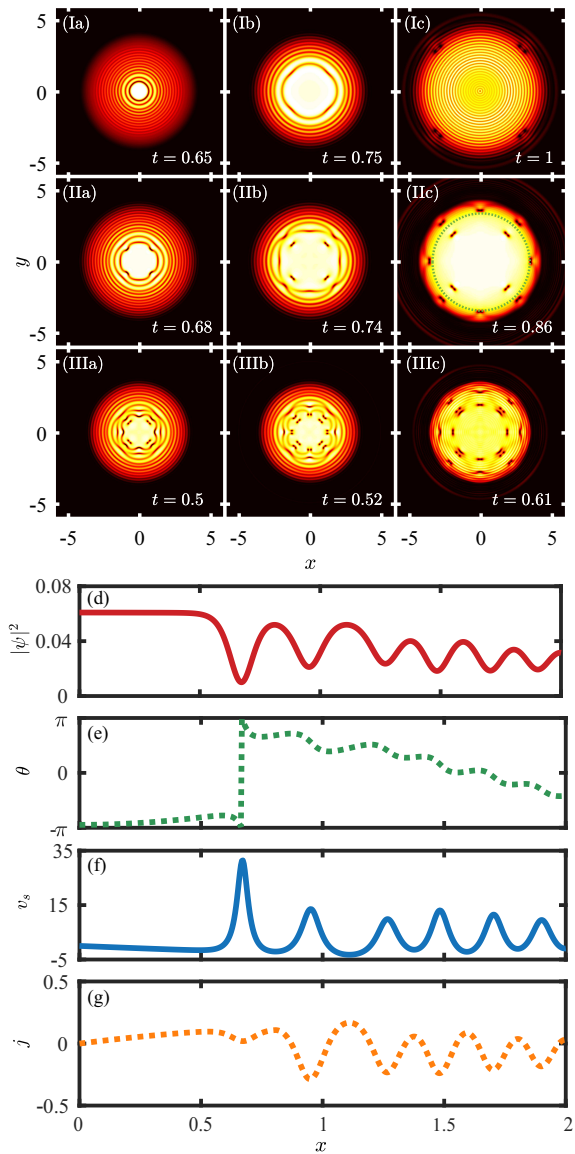}
\caption{(Color online) (I)-(III) Hot-scale plots of atom density (white = high). The parameters ($\alpha, c$) are (4, 6.4), (4, 7.8), and (5, 7.6), respectively. (d)-(g) The corresponding radial distributions of the density, phase, velocity and particle current of (Ia), respectively. Here $w=0.4$ and $b=2$.}
\label{Fig4}
\end{center}
\end{figure}

For the concave potential with $\alpha=4$, one dynamic RDS is excited in the system (see Fig.\,\ref{Fig4}\,(Ia)) with $c$ tuned from 1 to 6.4 suddenly (S-RDS regime). As shown in Fig.\,\ref{Fig4}\,(d) and (e), the RDS locates in the innermost shell of the concentric density rings, which is identified by both the density notch and a phase jump at $x=0.67$. All the other fringes are ring-shaped density wave oscillations where the condensate phase varies slowly and smoothly. The velocity of the condensate flow is proportional to the gradient of the phase, $v_s\propto\nabla{\theta}$, so a sharp phase variation means a large superfluid velocity (see Fig.\,\ref{Fig4}\,(f)). However, the low density induces rather flat and small particle current distribution $j(r)=|\psi(r)|^2v_s(r)$ near the RDS (see Fig.\,\ref{Fig4}\,(g)). As a result, the condensate in the regime surrounded by the RDS is free from density oscillations (see Figs.\,\ref{Fig4}\,(Ia), (Ib) and ({\sout{I}d})). Eventually, the RDS decays into 4 vortex-antivortex pairs via snake instability during its expansion towards the periphery of the condensate (see Figs.\,\ref{Fig4}\,(Ib) and (Ic)). The decay process is the same as the decay of the RDS originally imprinted in the harmonically trapped condensate described in Ref.\cite{PRL.90.120403}. The number of vortex pairs resulted in depends on the original depth of the input soliton for a given radius of the RDS, but always in multiples of four \cite{PRL.90.120403}. We note that the depth of a RDS is related to the phase modulation across the soliton, and for a given initial depth, the number of vortex pairs excited depends on the radius of the RDS. Larger radius may excite more pairs. 

\begin{figure*}[tbp]
\begin{center}
\includegraphics[angle=0,width=0.8\textwidth]{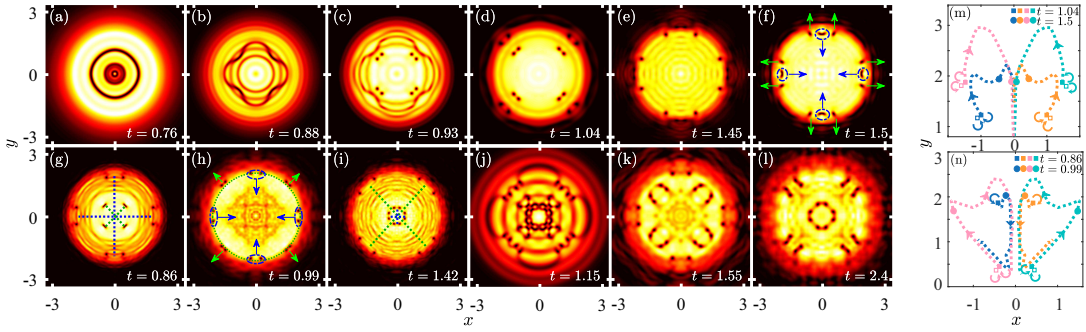}
\caption{(Color online) Hot-scale plots of atom density (white = high) {and typical vortex trajectories}. (a)-(f) D-RDS regime with $\alpha=0.6$ and $c=4.2$. (g)-(h) D-RDS regime with $\alpha=0.7$ and $c=7.2$. (i) D-RDS regime with $\alpha=0.7$ and $c=7$. (j)-(l) M-RDS regime with $\alpha=0.55$ and $c=4$. Here $w=1.2$ and $b=0.0002$. The blue and green arrows indicate the directions of the vortex pair movements. {(m) The vortex trajectories show the penetration of the vortices between the layers. (n) The vortex trajectories show the crisscross of the vortices between the layers. The solid squares and solid circles indicate the positions of the vortices at different times. The hollow squares are the positions of the counterparts of the vortices indicated by the solid squares.}}
\label{Fig5}
\end{center}
\end{figure*}
As shown in Fig.\,\ref{Fig4}\,(IIa), in the D-RDS regime, two concentric rings with low density appear at the center of the condensate cloud after $c$ is quenched from 1 to 7.8. These two rings are identified as RDSs due to the fact that their locations coincide well with phase jumps. The inner RDS is deeper and starts snaking earlier than the outer one, which is a precursor of splitting. Then it decays into 4 vortex pairs in $\times$-type configuration (see Fig.\,\ref{Fig4}\,(IIb)), similar to what happened in (Fig.\,\ref{Fig4}\,(Ic)). However, there are 4 additional lumps in the $+$-direction which move fast to the rim of the condensate. At the same time, the outer RDS begins to decay into 4 vortex pairs (see Fig.\,\ref{Fig4}\,(IIb)). Due to the existence of the outer RDS, the remaining lumps of the inner RDS will not annihilate in the rim of the condensate but turn into four more vortex pairs. The twelve vortex pairs then arrange themselves in three layers, forming a nested structure in $\times$-$+$-$\times$ configuration (see Fig.\,\ref{Fig4}\,(IIc)). {Layer structure of vortices has been identified in the collapse of ring dark solitons in a two-component BEC \cite{PRA.97.063607}.}

For $\alpha=5$, if we tune $c$ from 1 to 7.6, the dynamics of the system enter the M-RDS regime. In Fig.\,\ref{Fig4}\,(III), three RDSs are excited and then decay into 20 vortex pairs. The outermost RDS decays into 4 vortex pairs, while each of the other two splits into 8 vortex pairs. If there are $n$ RDSs excited during the quench, then they will decay into $4+ 8(n-1)$ vortex pairs. The initial distribution of these pairs right after the complete decay of the RDSs is the nested periodic $\times$-$+$-$\times$-$\cdots$ configuration as shown in Fig.\,\ref{Fig4}\,(IIIc).

For the convex potential, the dynamics of the system in the S-RDS regime is quite similar to that of the system trapped in the concave potential. The RDS decay into 4 vortex pairs. 
However, the dynamical excitation pattern has much richer structures when more than one RDS is exited in the system with a convex trap. We choose a set of typical parameters with $\alpha=0.6$ and $c=4.2$ to drive the system into D-RDS regime. With density and phase analysis, we can confirm that there are only two RDSs excited (see Fig.\,\ref{Fig5}\,(a)). The interesting fact is that the outer ring does not expand much, but bends heavily before the deformation of the inner ring  (Fig.\,\ref{Fig5}\,(b)).  The dynamics show different characteristics from the concave case. Instead of snaking, the inner RDS keeps the circular shape while expanding. After interacting with the outer RDS, 8 vortex pairs in the parallel $\times$-type configuration are excited in the condensate  (Figs.\,\ref{Fig5}\,(c-d)). Then the vortices and antivortices from these pairs move along the rings in two  layers and rearrange themselves into the parallel $+$-type configuration (Fig.\,\ref{Fig5}\,(e)). We note that this double layers of parallel vortex-antivortex pairs occurs in the energy higher than that of the nested vortex pairs.

Moreover, we found typical mechanisms of evolution between layers of vortex pairs, namely penetration and crisscross. For the parallel layers of vortex pairs, the quartets of vortex pairs in the outer layer can penetrate through the pairs of vortex and antivortex in the inner layer when they drift inward to the center of the condensate  (Fig.\,\ref{Fig5}\,(f)). Also, the the quartets of vortex pairs in the inner layer can penetrate through the vortex and antivortex pairs in the outer layer when they drift outward to the rim of the condensate. In these processes, a transient structure with an array of four linearly aligned vortex quadruple appears around the rim of the condensate. For the nested layers, the vortex pairs in two layers move towards each other along the $\times$- and $+$-directions (Figs.\,\ref{Fig5}\,(g)), respectively, and then generating crisscross trajectories and a necklace of vortex pairs around the rim of the condensate (Fig.\,\ref{Fig5}\,(h)). {The corresponding trajectories are shown in Figs.\,\ref{Fig5}\,(m) and (n).} The existence of the nested $+$-$\times$ and $\times$-$+$ configurations in the D-RDS regime is shown in Figs.\,\ref{Fig5}\,(g) and (i).

The configurations appearing during the dynamics in the M-RDS regime can be even more complicate  (Fig.\,\ref{Fig5}\,(j)). For $\alpha = 0.55$ and $c=4$, there are three RDSs excited. After a period of evolution, the system will oscillate between some typical configurations of 12 vortex pairs in three layers (Figs.\,\ref{Fig5}\,(k-l)). We find the hybrid configuration of  parallel and nested structures. We note that the recombination of vortices can also occur between different layers.

In both concave and convex potentials, after the decay of RDS(s), a transient stage appears during the evolution of vortex pairs, where a necklace of 8 vortices (vortex pairs) are set along a single ring (or an octagon structure). For 8 vortices (see Fig.\,\ref{Fig4}\,(Ic)), they will recombine themselves from $+$-type into $\times$-type vortex-antivortex pairs and vice versa. This is similar to the recombination of a vortex quadruple along the $x$- and $y$-direction \cite{SREP.6.29066}. However, for 8 vortex pairs (see Fig.\,\ref{Fig5}\,(h)), no recombination occurs between pairs from different layers (resulted from decay of different RDSs).

In conclusion, we have numerically studied interaction-quench dynamics and excitation patterns of 2D BEC confined in boxlike traps with different kinds of power-law boundaries. Relying on the strength of modulation, excitations with different number of RDS have been observed. For $\alpha \rightarrow 0$ or  $\alpha \rightarrow \infty $, it is easier to excite RDSs in the system. For $\alpha\in[0.5,0.6]$, the  parameter regions of S-RDS and D-RDS are extremely narrow, indicating that the condensate is very sensitive to the interaction quench. The decay behaviors of D-RDS and M-RDS  induced by quench dynamics are generally different {from} those of RDSs initially created by phase imprinting in the condensate. Moreover,  their decay dynamics strongly depend on the concavity and convexity of trap boundary. For a concave trap, the decay of dynamic RDSs obeys a universal rule. The inner RDS always splits into 8 vortex pairs, while the outmost RDS decays into 4 pairs. Typical configurations such as $\times$-type, $+$-type and their nested structures are found.  For a convex trap, richer structures of vortex pairs are developed, such as parallel and nested $\times$- and $+$-types, and their hybridization.  Concentric ring-shaped matter wave jets have been observed in the quench dynamics in concave traps, and the smaller $\alpha$ is, the easier to excite jets (see Figs.\,\ref{Fig4}\,(Ic-IIIc)).

\acknowledgments
This work is supported by the National Natural Science Foundation of China under grants Nos. 12175180, 11934015 and 11775178, the Major Basic Research Program of Natural Science of Shaanxi Province under grants Nos. 2017KCT-12 and 2017ZDJC-32. This research is also supported by The Double First-class University Construction Project of Northwest University.


\begin{thebibliography}{61}%
\makeatletter
\providecommand \@ifxundefined [1]{%
 \@ifx{#1\undefined}
}%
\providecommand \@ifnum [1]{%
 \ifnum #1\expandafter \@firstoftwo
 \else \expandafter \@secondoftwo
 \fi
}%
\providecommand \@ifx [1]{%
 \ifx #1\expandafter \@firstoftwo
 \else \expandafter \@secondoftwo
 \fi
}%
\providecommand \natexlab [1]{#1}%
\providecommand \enquote  [1]{``#1''}%
\providecommand \bibnamefont  [1]{#1}%
\providecommand \bibfnamefont [1]{#1}%
\providecommand \citenamefont [1]{#1}%
\providecommand \href@noop [0]{\@secondoftwo}%
\providecommand \href [0]{\begingroup \@sanitize@url \@href}%
\providecommand \@href[1]{\@@startlink{#1}\@@href}%
\providecommand \@@href[1]{\endgroup#1\@@endlink}%
\providecommand \@sanitize@url [0]{\catcode `\\12\catcode `\$12\catcode
  `\&12\catcode `\#12\catcode `\^12\catcode `\_12\catcode `\%12\relax}%
\providecommand \@@startlink[1]{}%
\providecommand \@@endlink[0]{}%
\providecommand \url  [0]{\begingroup\@sanitize@url \@url }%
\providecommand \@url [1]{\endgroup\@href {#1}{\urlprefix }}%
\providecommand \urlprefix  [0]{URL }%
\providecommand \Eprint [0]{\href }%
\providecommand \doibase [0]{http://dx.doi.org/}%
\providecommand \selectlanguage [0]{\@gobble}%
\providecommand \bibinfo  [0]{\@secondoftwo}%
\providecommand \bibfield  [0]{\@secondoftwo}%
\providecommand \translation [1]{[#1]}%
\providecommand \BibitemOpen [0]{}%
\providecommand \bibitemStop [0]{}%
\providecommand \bibitemNoStop [0]{.\EOS\space}%
\providecommand \EOS [0]{\spacefactor3000\relax}%
\providecommand \BibitemShut  [1]{\csname bibitem#1\endcsname}%
\let\auto@bib@innerbib\@empty
\bibitem [{\citenamefont {Eckardt}(2017)}]{Rev.Mod.Phys.89.011004}%
  \BibitemOpen
  \bibfield  {author} {\bibinfo {author} {\bibfnamefont {Andr\'e}\ \bibnamefont
  {Eckardt}},\ }\bibfield  {title} {\enquote {\bibinfo {title} {Colloquium:
  Atomic quantum gases in periodically driven optical lattices},}\ }\href
  {\doibase 10.1103/RevModPhys.89.011004} {\bibfield  {journal} {\bibinfo
  {journal} {Rev. Mod. Phys.}\ }\textbf {\bibinfo {volume} {89}},\ \bibinfo
  {pages} {011004} (\bibinfo {year} {2017})}\BibitemShut {NoStop}%
\bibitem [{\citenamefont {Dalibard}\ \emph {et~al.}(2011)\citenamefont
  {Dalibard}, \citenamefont {Gerbier}, \citenamefont
  {Juzeli\ifmmode~\bar{u}\else \={u}\fi{}nas},\ and\ \citenamefont
  {\"Ohberg}}]{Rev.Mod.Phys.83.1523}%
  \BibitemOpen
  \bibfield  {author} {\bibinfo {author} {\bibfnamefont {Jean}\ \bibnamefont
  {Dalibard}}, \bibinfo {author} {\bibfnamefont {Fabrice}\ \bibnamefont
  {Gerbier}}, \bibinfo {author} {\bibfnamefont {Gediminas}\ \bibnamefont
  {Juzeli\ifmmode~\bar{u}\else \={u}\fi{}nas}}, \ and\ \bibinfo {author}
  {\bibfnamefont {Patrik}\ \bibnamefont {\"Ohberg}},\ }\bibfield  {title}
  {\enquote {\bibinfo {title} {Colloquium: Artificial gauge potentials for
  neutral atoms},}\ }\href {\doibase 10.1103/RevModPhys.83.1523} {\bibfield
  {journal} {\bibinfo  {journal} {Rev. Mod. Phys.}\ }\textbf {\bibinfo {volume}
  {83}},\ \bibinfo {pages} {1523--1543} (\bibinfo {year} {2011})}\BibitemShut
  {NoStop}%
\bibitem [{\citenamefont {Jotzu}\ \emph {et~al.}(2014)\citenamefont {Jotzu},
  \citenamefont {Messer}, \citenamefont {Desbuquois}, \citenamefont {Lebrat},
  \citenamefont {Uehlinger}, \citenamefont {Greif},\ and\ \citenamefont
  {Esslinger}}]{Nature.515.237}%
  \BibitemOpen
  \bibfield  {author} {\bibinfo {author} {\bibfnamefont {Gregor}\ \bibnamefont
  {Jotzu}}, \bibinfo {author} {\bibfnamefont {Michael}\ \bibnamefont {Messer}},
  \bibinfo {author} {\bibfnamefont {R\'emi}\ \bibnamefont {Desbuquois}},
  \bibinfo {author} {\bibfnamefont {Martin}\ \bibnamefont {Lebrat}}, \bibinfo
  {author} {\bibfnamefont {Thomas}\ \bibnamefont {Uehlinger}}, \bibinfo
  {author} {\bibfnamefont {Daniel}\ \bibnamefont {Greif}}, \ and\ \bibinfo
  {author} {\bibfnamefont {Tilman.}\ \bibnamefont {Esslinger}},\ }\bibfield
  {title} {\enquote {\bibinfo {title} {Experimental realization of the
  topological {Haldane} model with ultracold fermions},}\ }\href {\doibase
  10.1038/nature13915} {\bibfield  {journal} {\bibinfo  {journal} {Nature}\
  }\textbf {\bibinfo {volume} {551}},\ \bibinfo {pages} {237--240} (\bibinfo
  {year} {2014})}\BibitemShut {NoStop}%
\bibitem [{\citenamefont {Mo}\ \emph {et~al.}(2020)\citenamefont {Mo},
  \citenamefont {Zhang},\ and\ \citenamefont {Wan}}]{CPL.37.060301}%
  \BibitemOpen
  \bibfield  {author} {\bibinfo {author} {\bibfnamefont {Lin-Han}\ \bibnamefont
  {Mo}}, \bibinfo {author} {\bibfnamefont {Qiu-Lan}\ \bibnamefont {Zhang}}, \
  and\ \bibinfo {author} {\bibfnamefont {Xin}\ \bibnamefont {Wan}},\ }\bibfield
   {title} {\enquote {\bibinfo {title} {Dynamics of the entanglement spectrum
  of the {H}aldane model under a sudden quench},}\ }\href {\doibase
  10.1088/0256-307x/37/6/060301} {\bibfield  {journal} {\bibinfo  {journal}
  {Chinese Physics Letters}\ }\textbf {\bibinfo {volume} {37}},\ \bibinfo
  {pages} {060301} (\bibinfo {year} {2020})}\BibitemShut {NoStop}%
\bibitem [{\citenamefont {Polkovnikov}\ \emph {et~al.}(2011)\citenamefont
  {Polkovnikov}, \citenamefont {Sengupta}, \citenamefont {Silva},\ and\
  \citenamefont {Vengalattore}}]{Rev.Mod.Phys.83.863}%
  \BibitemOpen
  \bibfield  {author} {\bibinfo {author} {\bibfnamefont {Anatoli}\ \bibnamefont
  {Polkovnikov}}, \bibinfo {author} {\bibfnamefont {Krishnendu}\ \bibnamefont
  {Sengupta}}, \bibinfo {author} {\bibfnamefont {Alessandro}\ \bibnamefont
  {Silva}}, \ and\ \bibinfo {author} {\bibfnamefont {Mukund}\ \bibnamefont
  {Vengalattore}},\ }\bibfield  {title} {\enquote {\bibinfo {title}
  {Colloquium: Nonequilibrium dynamics of closed interacting quantum
  systems},}\ }\href {\doibase 10.1103/RevModPhys.83.863} {\bibfield  {journal}
  {\bibinfo  {journal} {Rev. Mod. Phys.}\ }\textbf {\bibinfo {volume} {83}},\
  \bibinfo {pages} {863--883} (\bibinfo {year} {2011})}\BibitemShut {NoStop}%
\bibitem [{\citenamefont {Makotyn}\ \emph {et~al.}(2014)\citenamefont
  {Makotyn}, \citenamefont {Klauss}, \citenamefont {Goldberger}, \citenamefont
  {Cornell},\ and\ \citenamefont {Jin}}]{Nat.Phys.10.116}%
  \BibitemOpen
  \bibfield  {author} {\bibinfo {author} {\bibfnamefont {P.}~\bibnamefont
  {Makotyn}}, \bibinfo {author} {\bibfnamefont {C.~E.}\ \bibnamefont {Klauss}},
  \bibinfo {author} {\bibfnamefont {D.~L.}\ \bibnamefont {Goldberger}},
  \bibinfo {author} {\bibfnamefont {E.~A.}\ \bibnamefont {Cornell}}, \ and\
  \bibinfo {author} {\bibfnamefont {D.~S.}\ \bibnamefont {Jin}},\ }\bibfield
  {title} {\enquote {\bibinfo {title} {Universal dynamics of a degenerate
  unitary bose gas},}\ }\href {\doibase 10.1038/nphys2850} {\bibfield
  {journal} {\bibinfo  {journal} {Nature Physics}\ }\textbf {\bibinfo {volume}
  {10}},\ \bibinfo {pages} {116--119} (\bibinfo {year} {2014})}\BibitemShut
  {NoStop}%
\bibitem [{\citenamefont {Khaykovich}\ \emph {et~al.}(2002)\citenamefont
  {Khaykovich}, \citenamefont {Schreck}, \citenamefont {Ferrari}, \citenamefont
  {Bourdel}, \citenamefont {Cubizolles}, \citenamefont {Carr}, \citenamefont
  {Castin},\ and\ \citenamefont {Salomon}}]{science.1071021}%
  \BibitemOpen
  \bibfield  {author} {\bibinfo {author} {\bibfnamefont {L.}~\bibnamefont
  {Khaykovich}}, \bibinfo {author} {\bibfnamefont {F.}~\bibnamefont {Schreck}},
  \bibinfo {author} {\bibfnamefont {G.}~\bibnamefont {Ferrari}}, \bibinfo
  {author} {\bibfnamefont {T.}~\bibnamefont {Bourdel}}, \bibinfo {author}
  {\bibfnamefont {J.}~\bibnamefont {Cubizolles}}, \bibinfo {author}
  {\bibfnamefont {L.~D.}\ \bibnamefont {Carr}}, \bibinfo {author}
  {\bibfnamefont {Y.}~\bibnamefont {Castin}}, \ and\ \bibinfo {author}
  {\bibfnamefont {C.}~\bibnamefont {Salomon}},\ }\bibfield  {title} {\enquote
  {\bibinfo {title} {{Formation of a Matter-Wave Bright Soliton}},}\ }\href
  {\doibase 10.1126/science.1071021} {\bibfield  {journal} {\bibinfo  {journal}
  {Science}\ }\textbf {\bibinfo {volume} {296}},\ \bibinfo {pages} {1290--1293}
  (\bibinfo {year} {2002})}\BibitemShut {NoStop}%
\bibitem [{\citenamefont {Nguyen}\ \emph {et~al.}(2017)\citenamefont {Nguyen},
  \citenamefont {Luo},\ and\ \citenamefont {Hulet}}]{Nguyen422}%
  \BibitemOpen
  \bibfield  {author} {\bibinfo {author} {\bibfnamefont {Jason H~V}\
  \bibnamefont {Nguyen}}, \bibinfo {author} {\bibfnamefont {De}~\bibnamefont
  {Luo}}, \ and\ \bibinfo {author} {\bibfnamefont {Randall~G.}\ \bibnamefont
  {Hulet}},\ }\bibfield  {title} {\enquote {\bibinfo {title} {Formation of
  matter-wave soliton trains by modulational instability},}\ }\href {\doibase
  10.1126/science.aal3220} {\bibfield  {journal} {\bibinfo  {journal}
  {Science}\ }\textbf {\bibinfo {volume} {356}},\ \bibinfo {pages} {422--426}
  (\bibinfo {year} {2017})}\BibitemShut {NoStop}%
\bibitem [{\citenamefont {Carr}\ and\ \citenamefont
  {Brand}(2004)}]{PhysRevA.70.033607}%
  \BibitemOpen
  \bibfield  {author} {\bibinfo {author} {\bibfnamefont {L.~D.}\ \bibnamefont
  {Carr}}\ and\ \bibinfo {author} {\bibfnamefont {J.}~\bibnamefont {Brand}},\
  }\bibfield  {title} {\enquote {\bibinfo {title} {Pulsed atomic soliton
  laser},}\ }\href {\doibase 10.1103/PhysRevA.70.033607} {\bibfield  {journal}
  {\bibinfo  {journal} {Phys. Rev. A}\ }\textbf {\bibinfo {volume} {70}},\
  \bibinfo {pages} {033607} (\bibinfo {year} {2004})}\BibitemShut {NoStop}%
\bibitem [{\citenamefont {Feng-De}\ and\ \citenamefont
  {Jie-Fang}(2008)}]{CPL.25.2370}%
  \BibitemOpen
  \bibfield  {author} {\bibinfo {author} {\bibfnamefont {Zong}\ \bibnamefont
  {Feng-De}}\ and\ \bibinfo {author} {\bibfnamefont {Zhang}\ \bibnamefont
  {Jie-Fang}},\ }\bibfield  {title} {\enquote {\bibinfo {title} {Controlled
  manipulation with a {B}ose{\textendash}{E}instein condensates ${N}$-soliton
  train under the influence of harmonic and tilted periodic potentials},}\
  }\href {\doibase 10.1088/0256-307x/25/7/011} {\bibfield  {journal} {\bibinfo
  {journal} {Chinese Physics Letters}\ }\textbf {\bibinfo {volume} {25}},\
  \bibinfo {pages} {2370--2373} (\bibinfo {year} {2008})}\BibitemShut {NoStop}%
\bibitem [{\citenamefont {Liu}\ \emph {et~al.}(2009)\citenamefont {Liu},
  \citenamefont {Pu}, \citenamefont {Xiong}, \citenamefont {Liu},\ and\
  \citenamefont {Gong}}]{PhysRevA.79.013423}%
  \BibitemOpen
  \bibfield  {author} {\bibinfo {author} {\bibfnamefont {Xunxu}\ \bibnamefont
  {Liu}}, \bibinfo {author} {\bibfnamefont {Han}\ \bibnamefont {Pu}}, \bibinfo
  {author} {\bibfnamefont {Bo}~\bibnamefont {Xiong}}, \bibinfo {author}
  {\bibfnamefont {W.~M.}\ \bibnamefont {Liu}}, \ and\ \bibinfo {author}
  {\bibfnamefont {Jiangbin}\ \bibnamefont {Gong}},\ }\bibfield  {title}
  {\enquote {\bibinfo {title} {Formation and transformation of vector solitons
  in two-species {Bose-Einstein} condensates with a tunable interaction},}\
  }\href {\doibase 10.1103/PhysRevA.79.013423} {\bibfield  {journal} {\bibinfo
  {journal} {Phys. Rev. A}\ }\textbf {\bibinfo {volume} {79}},\ \bibinfo
  {pages} {013423} (\bibinfo {year} {2009})}\BibitemShut {NoStop}%
\bibitem [{\citenamefont {Abdullaev}\ \emph {et~al.}(2003)\citenamefont
  {Abdullaev}, \citenamefont {Caputo}, \citenamefont {Kraenkel},\ and\
  \citenamefont {Malomed}}]{PhysRevA.67.013605}%
  \BibitemOpen
  \bibfield  {author} {\bibinfo {author} {\bibfnamefont {Fatkhulla~Kh.}\
  \bibnamefont {Abdullaev}}, \bibinfo {author} {\bibfnamefont {Jean~Guy}\
  \bibnamefont {Caputo}}, \bibinfo {author} {\bibfnamefont {Robert~A.}\
  \bibnamefont {Kraenkel}}, \ and\ \bibinfo {author} {\bibfnamefont {Boris~A.}\
  \bibnamefont {Malomed}},\ }\bibfield  {title} {\enquote {\bibinfo {title}
  {Controlling collapse in {Bose-Einstein} condensates by temporal modulation
  of the scattering length},}\ }\href {\doibase 10.1103/PhysRevA.67.013605}
  {\bibfield  {journal} {\bibinfo  {journal} {Phys. Rev. A}\ }\textbf {\bibinfo
  {volume} {67}},\ \bibinfo {pages} {013605} (\bibinfo {year}
  {2003})}\BibitemShut {NoStop}%
\bibitem [{\citenamefont {Staliunas}\ \emph {et~al.}(2002)\citenamefont
  {Staliunas}, \citenamefont {Longhi},\ and\ \citenamefont
  {de~Valc\'arcel}}]{PhysRevLett.89.210406}%
  \BibitemOpen
  \bibfield  {author} {\bibinfo {author} {\bibfnamefont {Kestutis}\
  \bibnamefont {Staliunas}}, \bibinfo {author} {\bibfnamefont {Stefano}\
  \bibnamefont {Longhi}}, \ and\ \bibinfo {author} {\bibfnamefont
  {Germ\'an~J.}\ \bibnamefont {de~Valc\'arcel}},\ }\bibfield  {title} {\enquote
  {\bibinfo {title} {{Faraday Patterns in Bose-Einstein Condensates}},}\ }\href
  {\doibase 10.1103/PhysRevLett.89.210406} {\bibfield  {journal} {\bibinfo
  {journal} {Phys. Rev. Lett.}\ }\textbf {\bibinfo {volume} {89}},\ \bibinfo
  {pages} {210406} (\bibinfo {year} {2002})}\BibitemShut {NoStop}%
\bibitem [{\citenamefont {Pollack}\ \emph {et~al.}(2010)\citenamefont
  {Pollack}, \citenamefont {Dries}, \citenamefont {Hulet}, \citenamefont
  {Magalh\~aes}, \citenamefont {Henn}, \citenamefont {Ramos}, \citenamefont
  {Caracanhas},\ and\ \citenamefont {Bagnato}}]{PhysRevA.81.053627}%
  \BibitemOpen
  \bibfield  {author} {\bibinfo {author} {\bibfnamefont {S.~E.}\ \bibnamefont
  {Pollack}}, \bibinfo {author} {\bibfnamefont {D.}~\bibnamefont {Dries}},
  \bibinfo {author} {\bibfnamefont {R.~G.}\ \bibnamefont {Hulet}}, \bibinfo
  {author} {\bibfnamefont {K.~M.~F.}\ \bibnamefont {Magalh\~aes}}, \bibinfo
  {author} {\bibfnamefont {E.~A.~L.}\ \bibnamefont {Henn}}, \bibinfo {author}
  {\bibfnamefont {E.~R.~F.}\ \bibnamefont {Ramos}}, \bibinfo {author}
  {\bibfnamefont {M.~A.}\ \bibnamefont {Caracanhas}}, \ and\ \bibinfo {author}
  {\bibfnamefont {V.~S.}\ \bibnamefont {Bagnato}},\ }\bibfield  {title}
  {\enquote {\bibinfo {title} {Collective excitation of a {Bose-Einstein}
  condensate by modulation of the atomic scattering length},}\ }\href {\doibase
  10.1103/PhysRevA.81.053627} {\bibfield  {journal} {\bibinfo  {journal} {Phys.
  Rev. A}\ }\textbf {\bibinfo {volume} {81}},\ \bibinfo {pages} {053627}
  (\bibinfo {year} {2010})}\BibitemShut {NoStop}%
\bibitem [{\citenamefont {Vidanovi\ifmmode~\acute{c}\else \'{c}\fi{}}\ \emph
  {et~al.}(2011)\citenamefont {Vidanovi\ifmmode~\acute{c}\else \'{c}\fi{}},
  \citenamefont {Bala\ifmmode~\check{z}\else \v{z}\fi{}}, \citenamefont
  {Al-Jibbouri},\ and\ \citenamefont {Pelster}}]{PhysRevA.84.013618}%
  \BibitemOpen
  \bibfield  {author} {\bibinfo {author} {\bibfnamefont {Ivana}\ \bibnamefont
  {Vidanovi\ifmmode~\acute{c}\else \'{c}\fi{}}}, \bibinfo {author}
  {\bibfnamefont {Antun}\ \bibnamefont {Bala\ifmmode~\check{z}\else
  \v{z}\fi{}}}, \bibinfo {author} {\bibfnamefont {Hamid}\ \bibnamefont
  {Al-Jibbouri}}, \ and\ \bibinfo {author} {\bibfnamefont {Axel}\ \bibnamefont
  {Pelster}},\ }\bibfield  {title} {\enquote {\bibinfo {title} {Nonlinear
  {Bose-Einstein}-condensate dynamics induced by a harmonic modulation of the
  $s$-wave scattering length},}\ }\href {\doibase 10.1103/PhysRevA.84.013618}
  {\bibfield  {journal} {\bibinfo  {journal} {Phys. Rev. A}\ }\textbf {\bibinfo
  {volume} {84}},\ \bibinfo {pages} {013618} (\bibinfo {year}
  {2011})}\BibitemShut {NoStop}%
\bibitem [{\citenamefont {Clark}\ \emph {et~al.}(2017)\citenamefont {Clark},
  \citenamefont {Gaj}, \citenamefont {Feng},\ and\ \citenamefont
  {Chin}}]{Nature.551.356}%
  \BibitemOpen
  \bibfield  {author} {\bibinfo {author} {\bibfnamefont {Logan~W.}\
  \bibnamefont {Clark}}, \bibinfo {author} {\bibfnamefont {Anita}\ \bibnamefont
  {Gaj}}, \bibinfo {author} {\bibfnamefont {Lei}\ \bibnamefont {Feng}}, \ and\
  \bibinfo {author} {\bibfnamefont {Cheng.}\ \bibnamefont {Chin}},\ }\bibfield
  {title} {\enquote {\bibinfo {title} {Collective emission of matter-wave jets
  from driven {Bose-Einstein} condensates},}\ }\href {\doibase
  10.1038/nature24272} {\bibfield  {journal} {\bibinfo  {journal} {Nature}\
  }\textbf {\bibinfo {volume} {551}},\ \bibinfo {pages} {356--359} (\bibinfo
  {year} {2017})}\BibitemShut {NoStop}%
\bibitem [{\citenamefont {Fu}\ \emph {et~al.}(2018)\citenamefont {Fu},
  \citenamefont {Feng}, \citenamefont {Anderson}, \citenamefont {Clark},
  \citenamefont {Hu}, \citenamefont {Andrade}, \citenamefont {Chin},\ and\
  \citenamefont {Levin}}]{PhysRevLett.121.243001}%
  \BibitemOpen
  \bibfield  {author} {\bibinfo {author} {\bibfnamefont {Han}\ \bibnamefont
  {Fu}}, \bibinfo {author} {\bibfnamefont {Lei}\ \bibnamefont {Feng}}, \bibinfo
  {author} {\bibfnamefont {Brandon~M.}\ \bibnamefont {Anderson}}, \bibinfo
  {author} {\bibfnamefont {Logan~W.}\ \bibnamefont {Clark}}, \bibinfo {author}
  {\bibfnamefont {Jiazhong}\ \bibnamefont {Hu}}, \bibinfo {author}
  {\bibfnamefont {Jeffery~W.}\ \bibnamefont {Andrade}}, \bibinfo {author}
  {\bibfnamefont {Cheng}\ \bibnamefont {Chin}}, \ and\ \bibinfo {author}
  {\bibfnamefont {K.}~\bibnamefont {Levin}},\ }\bibfield  {title} {\enquote
  {\bibinfo {title} {{Density Waves and Jet Emission Asymmetry in Bose
  Fireworks}},}\ }\href {\doibase 10.1103/PhysRevLett.121.243001} {\bibfield
  {journal} {\bibinfo  {journal} {Phys. Rev. Lett.}\ }\textbf {\bibinfo
  {volume} {121}},\ \bibinfo {pages} {243001} (\bibinfo {year}
  {2018})}\BibitemShut {NoStop}%
\bibitem [{\citenamefont {Wu}\ and\ \citenamefont
  {Zhai}(2019)}]{PhysRevA.99.063624}%
  \BibitemOpen
  \bibfield  {author} {\bibinfo {author} {\bibfnamefont {Zhigang}\ \bibnamefont
  {Wu}}\ and\ \bibinfo {author} {\bibfnamefont {Hui}\ \bibnamefont {Zhai}},\
  }\bibfield  {title} {\enquote {\bibinfo {title} {Dynamics and density
  correlations in matter-wave jet emission of a driven condensate},}\ }\href
  {\doibase 10.1103/PhysRevA.99.063624} {\bibfield  {journal} {\bibinfo
  {journal} {Phys. Rev. A}\ }\textbf {\bibinfo {volume} {99}},\ \bibinfo
  {pages} {063624} (\bibinfo {year} {2019})}\BibitemShut {NoStop}%
\bibitem [{\citenamefont {Me\ifmmode \check{z}\else \v{z}\fi{}nar\ifmmode
  \check{s}\else \v{s}\fi{}i\ifmmode~\check{c}\else \v{c}\fi{}}\ \emph
  {et~al.}(2020)\citenamefont {Me\ifmmode \check{z}\else \v{z}\fi{}nar\ifmmode
  \check{s}\else \v{s}\fi{}i\ifmmode~\check{c}\else \v{c}\fi{}}, \citenamefont
  {\ifmmode~\check{Z}\else \v{Z}\fi{}itko}, \citenamefont {Arh}, \citenamefont
  {Gosar}, \citenamefont {Zupani\ifmmode~\check{c}\else \v{c}\fi{}},\ and\
  \citenamefont {Jegli\ifmmode~\check{c}\else
  \v{c}\fi{}}}]{PhysRevA.101.031601}%
  \BibitemOpen
  \bibfield  {author} {\bibinfo {author} {\bibfnamefont {Tadej}\ \bibnamefont
  {Me\ifmmode \check{z}\else \v{z}\fi{}nar\ifmmode \check{s}\else
  \v{s}\fi{}i\ifmmode~\check{c}\else \v{c}\fi{}}}, \bibinfo {author}
  {\bibfnamefont {Rok}\ \bibnamefont {\ifmmode~\check{Z}\else \v{Z}\fi{}itko}},
  \bibinfo {author} {\bibfnamefont {Tina}\ \bibnamefont {Arh}}, \bibinfo
  {author} {\bibfnamefont {Katja}\ \bibnamefont {Gosar}}, \bibinfo {author}
  {\bibfnamefont {Erik}\ \bibnamefont {Zupani\ifmmode~\check{c}\else
  \v{c}\fi{}}}, \ and\ \bibinfo {author} {\bibfnamefont {Peter}\ \bibnamefont
  {Jegli\ifmmode~\check{c}\else \v{c}\fi{}}},\ }\bibfield  {title} {\enquote
  {\bibinfo {title} {Emission of correlated jets from a driven matter-wave
  soliton in a quasi-one-dimensional geometry},}\ }\href {\doibase
  10.1103/PhysRevA.101.031601} {\bibfield  {journal} {\bibinfo  {journal}
  {Phys. Rev. A}\ }\textbf {\bibinfo {volume} {101}},\ \bibinfo {pages}
  {031601} (\bibinfo {year} {2020})}\BibitemShut {NoStop}%
\bibitem [{\citenamefont {Kivshar}\ and\ \citenamefont
  {Luther-Davies}(1998)}]{Phys.Rep.298.81}%
  \BibitemOpen
  \bibfield  {author} {\bibinfo {author} {\bibfnamefont {Yuri~S.}\ \bibnamefont
  {Kivshar}}\ and\ \bibinfo {author} {\bibfnamefont {Barry}\ \bibnamefont
  {Luther-Davies}},\ }\bibfield  {title} {\enquote {\bibinfo {title} {Dark
  optical solitons: physics and applications},}\ }\href {\doibase
  https://doi.org/10.1016/S0370-1573(97)00073-2} {\bibfield  {journal}
  {\bibinfo  {journal} {Physics Reports}\ }\textbf {\bibinfo {volume} {298}},\
  \bibinfo {pages} {81--197} (\bibinfo {year} {1998})}\BibitemShut {NoStop}%
\bibitem [{\citenamefont {Kivshar}\ \emph {et~al.}(1998)\citenamefont
  {Kivshar}, \citenamefont {Christou}, \citenamefont {Tikhonenko},
  \citenamefont {Luther-Davies},\ and\ \citenamefont
  {Pismen}}]{Opt.Commun.152.198}%
  \BibitemOpen
  \bibfield  {author} {\bibinfo {author} {\bibfnamefont {Yuri~S.}\ \bibnamefont
  {Kivshar}}, \bibinfo {author} {\bibfnamefont {Jason}\ \bibnamefont
  {Christou}}, \bibinfo {author} {\bibfnamefont {Vladimir}\ \bibnamefont
  {Tikhonenko}}, \bibinfo {author} {\bibfnamefont {Barry}\ \bibnamefont
  {Luther-Davies}}, \ and\ \bibinfo {author} {\bibfnamefont {Len~M.}\
  \bibnamefont {Pismen}},\ }\bibfield  {title} {\enquote {\bibinfo {title}
  {Dynamics of optical vortex solitons},}\ }\href {\doibase
  https://doi.org/10.1016/S0030-4018(98)00149-7} {\bibfield  {journal}
  {\bibinfo  {journal} {Optics Communications}\ }\textbf {\bibinfo {volume}
  {152}},\ \bibinfo {pages} {198--206} (\bibinfo {year} {1998})}\BibitemShut
  {NoStop}%
\bibitem [{\citenamefont {Kivotides}\ \emph {et~al.}(2000)\citenamefont
  {Kivotides}, \citenamefont {Barenghi},\ and\ \citenamefont
  {Samuels}}]{Science.290.777}%
  \BibitemOpen
  \bibfield  {author} {\bibinfo {author} {\bibfnamefont {Demosthenes}\
  \bibnamefont {Kivotides}}, \bibinfo {author} {\bibfnamefont {Carlo~F.}\
  \bibnamefont {Barenghi}}, \ and\ \bibinfo {author} {\bibfnamefont {David~C.}\
  \bibnamefont {Samuels}},\ }\bibfield  {title} {\enquote {\bibinfo {title}
  {{Triple Vortex Ring Structure in Superfluid Helium II}},}\ }\href {\doibase
  10.1126/science.290.5492.777} {\bibfield  {journal} {\bibinfo  {journal}
  {Science}\ }\textbf {\bibinfo {volume} {290}},\ \bibinfo {pages} {777--779}
  (\bibinfo {year} {2000})}\BibitemShut {NoStop}%
\bibitem [{\citenamefont {Bai}\ \emph {et~al.}(2020)\citenamefont {Bai},
  \citenamefont {Yang},\ and\ \citenamefont {Liu}}]{PhysRevA.102.063318}%
  \BibitemOpen
  \bibfield  {author} {\bibinfo {author} {\bibfnamefont {Wen-Kai}\ \bibnamefont
  {Bai}}, \bibinfo {author} {\bibfnamefont {Tao}\ \bibnamefont {Yang}}, \ and\
  \bibinfo {author} {\bibfnamefont {Wu-Ming}\ \bibnamefont {Liu}},\ }\bibfield
  {title} {\enquote {\bibinfo {title} {Topological transition from superfluid
  vortex rings to isolated knots and links},}\ }\href {\doibase
  10.1103/PhysRevA.102.063318} {\bibfield  {journal} {\bibinfo  {journal}
  {Phys. Rev. A}\ }\textbf {\bibinfo {volume} {102}},\ \bibinfo {pages}
  {063318} (\bibinfo {year} {2020})}\BibitemShut {NoStop}%
\bibitem [{\citenamefont {Kopnin}(2002)}]{Kopnin_2002}%
  \BibitemOpen
  \bibfield  {author} {\bibinfo {author} {\bibfnamefont {N~B}\ \bibnamefont
  {Kopnin}},\ }\bibfield  {title} {\enquote {\bibinfo {title} {Vortex dynamics
  and mutual friction in superconductors and fermi superfluids},}\ }\href@noop
  {} {\bibfield  {journal} {\bibinfo  {journal} {Reports on Progress in
  Physics}\ }\textbf {\bibinfo {volume} {65}},\ \bibinfo {pages} {1633--1678}
  (\bibinfo {year} {2002})}\BibitemShut {NoStop}%
\bibitem [{\citenamefont {Burger}\ \emph {et~al.}(1999)\citenamefont {Burger},
  \citenamefont {Bongs}, \citenamefont {Dettmer}, \citenamefont {Ertmer},
  \citenamefont {Sengstock}, \citenamefont {Sanpera}, \citenamefont
  {Shlyapnikov},\ and\ \citenamefont {Lewenstein}}]{PRL.83.5198}%
  \BibitemOpen
  \bibfield  {author} {\bibinfo {author} {\bibfnamefont {S.}~\bibnamefont
  {Burger}}, \bibinfo {author} {\bibfnamefont {K.}~\bibnamefont {Bongs}},
  \bibinfo {author} {\bibfnamefont {S.}~\bibnamefont {Dettmer}}, \bibinfo
  {author} {\bibfnamefont {W.}~\bibnamefont {Ertmer}}, \bibinfo {author}
  {\bibfnamefont {K.}~\bibnamefont {Sengstock}}, \bibinfo {author}
  {\bibfnamefont {A.}~\bibnamefont {Sanpera}}, \bibinfo {author} {\bibfnamefont
  {G.~V.}\ \bibnamefont {Shlyapnikov}}, \ and\ \bibinfo {author} {\bibfnamefont
  {M.}~\bibnamefont {Lewenstein}},\ }\bibfield  {title} {\enquote {\bibinfo
  {title} {{Dark Solitons in Bose-Einstein Condensates}},}\ }\href {\doibase
  10.1103/PhysRevLett.83.5198} {\bibfield  {journal} {\bibinfo  {journal}
  {Phys. Rev. Lett.}\ }\textbf {\bibinfo {volume} {83}},\ \bibinfo {pages}
  {5198--5201} (\bibinfo {year} {1999})}\BibitemShut {NoStop}%
\bibitem [{\citenamefont {Denschlag}\ \emph {et~al.}(2000)\citenamefont
  {Denschlag}, \citenamefont {Simsarian}, \citenamefont {Feder}, \citenamefont
  {Clark}, \citenamefont {Collins}, \citenamefont {Cubizolles}, \citenamefont
  {Deng}, \citenamefont {Hagley}, \citenamefont {Helmerson}, \citenamefont
  {Reinhardt}, \citenamefont {Rolston}, \citenamefont {Schneider},\ and\
  \citenamefont {Phillips}}]{Science.287.97}%
  \BibitemOpen
  \bibfield  {author} {\bibinfo {author} {\bibfnamefont {J.}~\bibnamefont
  {Denschlag}}, \bibinfo {author} {\bibfnamefont {J.~E.}\ \bibnamefont
  {Simsarian}}, \bibinfo {author} {\bibfnamefont {D.~L.}\ \bibnamefont
  {Feder}}, \bibinfo {author} {\bibfnamefont {Charles~W.}\ \bibnamefont
  {Clark}}, \bibinfo {author} {\bibfnamefont {L.~A.}\ \bibnamefont {Collins}},
  \bibinfo {author} {\bibfnamefont {J.}~\bibnamefont {Cubizolles}}, \bibinfo
  {author} {\bibfnamefont {L.}~\bibnamefont {Deng}}, \bibinfo {author}
  {\bibfnamefont {E.~W.}\ \bibnamefont {Hagley}}, \bibinfo {author}
  {\bibfnamefont {K.}~\bibnamefont {Helmerson}}, \bibinfo {author}
  {\bibfnamefont {W.~P.}\ \bibnamefont {Reinhardt}}, \bibinfo {author}
  {\bibfnamefont {S.~L.}\ \bibnamefont {Rolston}}, \bibinfo {author}
  {\bibfnamefont {B.~I.}\ \bibnamefont {Schneider}}, \ and\ \bibinfo {author}
  {\bibfnamefont {W.~D.}\ \bibnamefont {Phillips}},\ }\bibfield  {title}
  {\enquote {\bibinfo {title} {{Generating Solitons by Phase Engineering of a
  Bose-Einstein Condensate}},}\ }\href {\doibase 10.1126/science.287.5450.97}
  {\bibfield  {journal} {\bibinfo  {journal} {Science}\ }\textbf {\bibinfo
  {volume} {287}},\ \bibinfo {pages} {97--101} (\bibinfo {year}
  {2000})}\BibitemShut {NoStop}%
\bibitem [{\citenamefont {Fritsch}\ \emph {et~al.}(2020)\citenamefont
  {Fritsch}, \citenamefont {Lu}, \citenamefont {Reid}, \citenamefont
  {Pi\~neiro},\ and\ \citenamefont {Spielman}}]{PhysRevA.101.053629}%
  \BibitemOpen
  \bibfield  {author} {\bibinfo {author} {\bibfnamefont {A.~R.}\ \bibnamefont
  {Fritsch}}, \bibinfo {author} {\bibfnamefont {Mingwu}\ \bibnamefont {Lu}},
  \bibinfo {author} {\bibfnamefont {G.~H.}\ \bibnamefont {Reid}}, \bibinfo
  {author} {\bibfnamefont {A.~M.}\ \bibnamefont {Pi\~neiro}}, \ and\ \bibinfo
  {author} {\bibfnamefont {I.~B.}\ \bibnamefont {Spielman}},\ }\bibfield
  {title} {\enquote {\bibinfo {title} {Creating solitons with controllable and
  near-zero velocity in {Bose-Einstein} condensates},}\ }\href {\doibase
  10.1103/PhysRevA.101.053629} {\bibfield  {journal} {\bibinfo  {journal}
  {Phys. Rev. A}\ }\textbf {\bibinfo {volume} {101}},\ \bibinfo {pages}
  {053629} (\bibinfo {year} {2020})}\BibitemShut {NoStop}%
\bibitem [{\citenamefont {Dutton}\ \emph {et~al.}(2001)\citenamefont {Dutton},
  \citenamefont {Budde}, \citenamefont {Slowe},\ and\ \citenamefont
  {Hau}}]{Science.293.663}%
  \BibitemOpen
  \bibfield  {author} {\bibinfo {author} {\bibfnamefont {Zachary}\ \bibnamefont
  {Dutton}}, \bibinfo {author} {\bibfnamefont {Michael}\ \bibnamefont {Budde}},
  \bibinfo {author} {\bibfnamefont {Christopher}\ \bibnamefont {Slowe}}, \ and\
  \bibinfo {author} {\bibfnamefont {Lene~Vestergaard}\ \bibnamefont {Hau}},\
  }\bibfield  {title} {\enquote {\bibinfo {title} {{Observation of Quantum
  Shock Waves Created with Ultra-Compressed Slow Light Pulses in a
  Bose-Einstein Condensate}},}\ }\href {\doibase 10.1126/science.1062527}
  {\bibfield  {journal} {\bibinfo  {journal} {Science}\ }\textbf {\bibinfo
  {volume} {293}},\ \bibinfo {pages} {663--668} (\bibinfo {year}
  {2001})}\BibitemShut {NoStop}%
\bibitem [{\citenamefont {Weller}\ \emph {et~al.}(2008)\citenamefont {Weller},
  \citenamefont {Ronzheimer}, \citenamefont {Gross}, \citenamefont {Esteve},
  \citenamefont {Oberthaler}, \citenamefont {Frantzeskakis}, \citenamefont
  {Theocharis},\ and\ \citenamefont {Kevrekidis}}]{PhysRevLett.101.130401}%
  \BibitemOpen
  \bibfield  {author} {\bibinfo {author} {\bibfnamefont {A.}~\bibnamefont
  {Weller}}, \bibinfo {author} {\bibfnamefont {J.~P.}\ \bibnamefont
  {Ronzheimer}}, \bibinfo {author} {\bibfnamefont {C.}~\bibnamefont {Gross}},
  \bibinfo {author} {\bibfnamefont {J.}~\bibnamefont {Esteve}}, \bibinfo
  {author} {\bibfnamefont {M.~K.}\ \bibnamefont {Oberthaler}}, \bibinfo
  {author} {\bibfnamefont {D.~J.}\ \bibnamefont {Frantzeskakis}}, \bibinfo
  {author} {\bibfnamefont {G.}~\bibnamefont {Theocharis}}, \ and\ \bibinfo
  {author} {\bibfnamefont {P.~G.}\ \bibnamefont {Kevrekidis}},\ }\bibfield
  {title} {\enquote {\bibinfo {title} {{Experimental Observation of Oscillating
  and Interacting Matter Wave Dark Solitons}},}\ }\href {\doibase
  10.1103/PhysRevLett.101.130401} {\bibfield  {journal} {\bibinfo  {journal}
  {Phys. Rev. Lett.}\ }\textbf {\bibinfo {volume} {101}},\ \bibinfo {pages}
  {130401} (\bibinfo {year} {2008})}\BibitemShut {NoStop}%
\bibitem [{\citenamefont {Yang}\ \emph {et~al.}(2013)\citenamefont {Yang},
  \citenamefont {Xiong},\ and\ \citenamefont {Benedict}}]{PhysRevA.87.023603}%
  \BibitemOpen
  \bibfield  {author} {\bibinfo {author} {\bibfnamefont {T.}~\bibnamefont
  {Yang}}, \bibinfo {author} {\bibfnamefont {B.}~\bibnamefont {Xiong}}, \ and\
  \bibinfo {author} {\bibfnamefont {Keith~A.}\ \bibnamefont {Benedict}},\
  }\bibfield  {title} {\enquote {\bibinfo {title} {Dynamical excitations in the
  collision of two-dimensional {Bose-Einstein} condensates},}\ }\href {\doibase
  10.1103/PhysRevA.87.023603} {\bibfield  {journal} {\bibinfo  {journal} {Phys.
  Rev. A}\ }\textbf {\bibinfo {volume} {87}},\ \bibinfo {pages} {023603}
  (\bibinfo {year} {2013})}\BibitemShut {NoStop}%
\bibitem [{\citenamefont {Xiong}\ \emph {et~al.}(2013)\citenamefont {Xiong},
  \citenamefont {Yang},\ and\ \citenamefont {Benedict}}]{PhysRevA.88.043602}%
  \BibitemOpen
  \bibfield  {author} {\bibinfo {author} {\bibfnamefont {Bo}~\bibnamefont
  {Xiong}}, \bibinfo {author} {\bibfnamefont {Tao}\ \bibnamefont {Yang}}, \
  and\ \bibinfo {author} {\bibfnamefont {Keith~A.}\ \bibnamefont {Benedict}},\
  }\bibfield  {title} {\enquote {\bibinfo {title} {Distortion of interference
  fringes and the resulting vortex production of merging {Bose-Einstein}
  condensates},}\ }\href {\doibase 10.1103/PhysRevA.88.043602} {\bibfield
  {journal} {\bibinfo  {journal} {Phys. Rev. A}\ }\textbf {\bibinfo {volume}
  {88}},\ \bibinfo {pages} {043602} (\bibinfo {year} {2013})}\BibitemShut
  {NoStop}%
\bibitem [{\citenamefont {Halperin}\ and\ \citenamefont
  {Bohn}(2020)}]{PhysRevResearch.2.043256}%
  \BibitemOpen
  \bibfield  {author} {\bibinfo {author} {\bibfnamefont {E.~J.}\ \bibnamefont
  {Halperin}}\ and\ \bibinfo {author} {\bibfnamefont {J.~L.}\ \bibnamefont
  {Bohn}},\ }\bibfield  {title} {\enquote {\bibinfo {title} {Quench-produced
  solitons in a box-trapped {Bose-Einstein} condensate},}\ }\href {\doibase
  10.1103/PhysRevResearch.2.043256} {\bibfield  {journal} {\bibinfo  {journal}
  {Phys. Rev. Research}\ }\textbf {\bibinfo {volume} {2}},\ \bibinfo {pages}
  {043256} (\bibinfo {year} {2020})}\BibitemShut {NoStop}%
\bibitem [{\citenamefont {Jia}\ and\ \citenamefont
  {Liang}(2020)}]{CPL.37.040502}%
  \BibitemOpen
  \bibfield  {author} {\bibinfo {author} {\bibfnamefont {Chun-Yu}\ \bibnamefont
  {Jia}}\ and\ \bibinfo {author} {\bibfnamefont {Zhao-Xin}\ \bibnamefont
  {Liang}},\ }\bibfield  {title} {\enquote {\bibinfo {title} {Dark soliton of
  polariton condensates under nonresonant {PT}-symmetric pumping},}\ }\href
  {\doibase 10.1088/0256-307x/37/4/040502} {\bibfield  {journal} {\bibinfo
  {journal} {Chinese Physics Letters}\ }\textbf {\bibinfo {volume} {37}},\
  \bibinfo {pages} {040502} (\bibinfo {year} {2020})}\BibitemShut {NoStop}%
\bibitem [{\citenamefont {Wen}\ \emph {et~al.}(2016)\citenamefont {Wen},
  \citenamefont {Sun}, \citenamefont {Chen}, \citenamefont {Wang},
  \citenamefont {Hu}, \citenamefont {Chen}, \citenamefont {Liu}, \citenamefont
  {Juzeli\ifmmode~\bar{u}\else \={u}\fi{}nas}, \citenamefont {Malomed},\ and\
  \citenamefont {Ji}}]{PhysRevA.94.061602}%
  \BibitemOpen
  \bibfield  {author} {\bibinfo {author} {\bibfnamefont {Lin}\ \bibnamefont
  {Wen}}, \bibinfo {author} {\bibfnamefont {Q.}~\bibnamefont {Sun}}, \bibinfo
  {author} {\bibfnamefont {Yu}~\bibnamefont {Chen}}, \bibinfo {author}
  {\bibfnamefont {Deng-Shan}\ \bibnamefont {Wang}}, \bibinfo {author}
  {\bibfnamefont {J.}~\bibnamefont {Hu}}, \bibinfo {author} {\bibfnamefont
  {H.}~\bibnamefont {Chen}}, \bibinfo {author} {\bibfnamefont {W.-M.}\
  \bibnamefont {Liu}}, \bibinfo {author} {\bibfnamefont {G.}~\bibnamefont
  {Juzeli\ifmmode~\bar{u}\else \={u}\fi{}nas}}, \bibinfo {author}
  {\bibfnamefont {Boris~A.}\ \bibnamefont {Malomed}}, \ and\ \bibinfo {author}
  {\bibfnamefont {An-Chun}\ \bibnamefont {Ji}},\ }\bibfield  {title} {\enquote
  {\bibinfo {title} {Motion of solitons in one-dimensional spin-orbit-coupled
  {B}ose-{E}instein condensates},}\ }\href {\doibase
  10.1103/PhysRevA.94.061602} {\bibfield  {journal} {\bibinfo  {journal} {Phys.
  Rev. A}\ }\textbf {\bibinfo {volume} {94}},\ \bibinfo {pages} {061602}
  (\bibinfo {year} {2016})}\BibitemShut {NoStop}%
\bibitem [{\citenamefont {Kivshar}\ and\ \citenamefont
  {Pelinovsky}(2000)}]{Phys.Rep.331.117}%
  \BibitemOpen
  \bibfield  {author} {\bibinfo {author} {\bibfnamefont {Yuri~S.}\ \bibnamefont
  {Kivshar}}\ and\ \bibinfo {author} {\bibfnamefont {Dmitry~E.}\ \bibnamefont
  {Pelinovsky}},\ }\bibfield  {title} {\enquote {\bibinfo {title}
  {Self-focusing and transverse instabilities of solitary waves},}\ }\href
  {\doibase https://doi.org/10.1016/S0370-1573(99)00106-4} {\bibfield
  {journal} {\bibinfo  {journal} {Physics Reports}\ }\textbf {\bibinfo {volume}
  {331}},\ \bibinfo {pages} {117--195} (\bibinfo {year} {2000})}\BibitemShut
  {NoStop}%
\bibitem [{\citenamefont {Theocharis}\ \emph {et~al.}(2003)\citenamefont
  {Theocharis}, \citenamefont {Frantzeskakis}, \citenamefont {Kevrekidis},
  \citenamefont {Malomed},\ and\ \citenamefont {Kivshar}}]{PRL.90.120403}%
  \BibitemOpen
  \bibfield  {author} {\bibinfo {author} {\bibfnamefont {G.}~\bibnamefont
  {Theocharis}}, \bibinfo {author} {\bibfnamefont {D.~J.}\ \bibnamefont
  {Frantzeskakis}}, \bibinfo {author} {\bibfnamefont {P.~G.}\ \bibnamefont
  {Kevrekidis}}, \bibinfo {author} {\bibfnamefont {B.~A.}\ \bibnamefont
  {Malomed}}, \ and\ \bibinfo {author} {\bibfnamefont {Yuri~S.}\ \bibnamefont
  {Kivshar}},\ }\bibfield  {title} {\enquote {\bibinfo {title} {{Ring Dark
  Solitons and Vortex Necklaces in Bose-Einstein Condensates}},}\ }\href
  {\doibase 10.1103/PhysRevLett.90.120403} {\bibfield  {journal} {\bibinfo
  {journal} {Phys. Rev. Lett.}\ }\textbf {\bibinfo {volume} {90}},\ \bibinfo
  {pages} {120403} (\bibinfo {year} {2003})}\BibitemShut {NoStop}%
\bibitem [{\citenamefont {Wang}\ \emph {et~al.}(2015)\citenamefont {Wang},
  \citenamefont {Kevrekidis}, \citenamefont {Carretero-Gonz\'alez},
  \citenamefont {Frantzeskakis}, \citenamefont {Kaper},\ and\ \citenamefont
  {Ma}}]{PhysRevA.92.033611}%
  \BibitemOpen
  \bibfield  {author} {\bibinfo {author} {\bibfnamefont {Wenlong}\ \bibnamefont
  {Wang}}, \bibinfo {author} {\bibfnamefont {P.~G.}\ \bibnamefont
  {Kevrekidis}}, \bibinfo {author} {\bibfnamefont {R.}~\bibnamefont
  {Carretero-Gonz\'alez}}, \bibinfo {author} {\bibfnamefont {D.~J.}\
  \bibnamefont {Frantzeskakis}}, \bibinfo {author} {\bibfnamefont {Tasso~J.}\
  \bibnamefont {Kaper}}, \ and\ \bibinfo {author} {\bibfnamefont {Manjun}\
  \bibnamefont {Ma}},\ }\bibfield  {title} {\enquote {\bibinfo {title}
  {Stabilization of ring dark solitons in {Bose-Einstein} condensates},}\
  }\href {\doibase 10.1103/PhysRevA.92.033611} {\bibfield  {journal} {\bibinfo
  {journal} {Phys. Rev. A}\ }\textbf {\bibinfo {volume} {92}},\ \bibinfo
  {pages} {033611} (\bibinfo {year} {2015})}\BibitemShut {NoStop}%
\bibitem [{\citenamefont {Song}\ \emph {et~al.}(2012)\citenamefont {Song},
  \citenamefont {Wang}, \citenamefont {Wang},\ and\ \citenamefont
  {Liu}}]{PhysRevA.85.063617}%
  \BibitemOpen
  \bibfield  {author} {\bibinfo {author} {\bibfnamefont {Shu-Wei}\ \bibnamefont
  {Song}}, \bibinfo {author} {\bibfnamefont {Deng-Shan}\ \bibnamefont {Wang}},
  \bibinfo {author} {\bibfnamefont {Hanquan}\ \bibnamefont {Wang}}, \ and\
  \bibinfo {author} {\bibfnamefont {W.~M.}\ \bibnamefont {Liu}},\ }\bibfield
  {title} {\enquote {\bibinfo {title} {Generation of ring dark solitons by
  phase engineering and their oscillations in spin-1 {Bose-Einstein}
  condensates},}\ }\href {\doibase 10.1103/PhysRevA.85.063617} {\bibfield
  {journal} {\bibinfo  {journal} {Phys. Rev. A}\ }\textbf {\bibinfo {volume}
  {85}},\ \bibinfo {pages} {063617} (\bibinfo {year} {2012})}\BibitemShut
  {NoStop}%
\bibitem [{\citenamefont {Hu}\ \emph {et~al.}(2009)\citenamefont {Hu},
  \citenamefont {Zhang}, \citenamefont {Zhao}, \citenamefont {Luo},\ and\
  \citenamefont {Liu}}]{PhysRevA.79.023619}%
  \BibitemOpen
  \bibfield  {author} {\bibinfo {author} {\bibfnamefont {Xing-Hua}\
  \bibnamefont {Hu}}, \bibinfo {author} {\bibfnamefont {Xiao-Fei}\ \bibnamefont
  {Zhang}}, \bibinfo {author} {\bibfnamefont {Dun}\ \bibnamefont {Zhao}},
  \bibinfo {author} {\bibfnamefont {Hong-Gang}\ \bibnamefont {Luo}}, \ and\
  \bibinfo {author} {\bibfnamefont {W.~M.}\ \bibnamefont {Liu}},\ }\bibfield
  {title} {\enquote {\bibinfo {title} {Dynamics and modulation of ring dark
  solitons in two-dimensional {Bose-Einstein} condensates with tunable
  interaction},}\ }\href {\doibase 10.1103/PhysRevA.79.023619} {\bibfield
  {journal} {\bibinfo  {journal} {Phys. Rev. A}\ }\textbf {\bibinfo {volume}
  {79}},\ \bibinfo {pages} {023619} (\bibinfo {year} {2009})}\BibitemShut
  {NoStop}%
\bibitem [{\citenamefont {Wang}\ \emph
  {et~al.}(2021{\natexlab{a}})\citenamefont {Wang}, \citenamefont
  {Kolokolnikov}, \citenamefont {Frantzeskakis}, \citenamefont
  {Carretero-Gonz\'alez},\ and\ \citenamefont
  {Kevrekidis}}]{PhysRevA.104.023314}%
  \BibitemOpen
  \bibfield  {author} {\bibinfo {author} {\bibfnamefont {Wenlong}\ \bibnamefont
  {Wang}}, \bibinfo {author} {\bibfnamefont {Theodore}\ \bibnamefont
  {Kolokolnikov}}, \bibinfo {author} {\bibfnamefont {D.~J.}\ \bibnamefont
  {Frantzeskakis}}, \bibinfo {author} {\bibfnamefont {R.}~\bibnamefont
  {Carretero-Gonz\'alez}}, \ and\ \bibinfo {author} {\bibfnamefont {P.~G.}\
  \bibnamefont {Kevrekidis}},\ }\bibfield  {title} {\enquote {\bibinfo {title}
  {Pairwise interactions of ring dark solitons with vortices and other rings:
  Stationary states, stability features, and nonlinear dynamics},}\ }\href
  {\doibase 10.1103/PhysRevA.104.023314} {\bibfield  {journal} {\bibinfo
  {journal} {Phys. Rev. A}\ }\textbf {\bibinfo {volume} {104}},\ \bibinfo
  {pages} {023314} (\bibinfo {year} {2021}{\natexlab{a}})}\BibitemShut
  {NoStop}%
\bibitem [{\citenamefont {Yang}\ \emph {et~al.}(2007)\citenamefont {Yang},
  \citenamefont {Wu}, \citenamefont {Zhang}, \citenamefont {Feng},
  \citenamefont {Guo}, \citenamefont {Wen},\ and\ \citenamefont
  {Yu}}]{PhysRevA.76.063606}%
  \BibitemOpen
  \bibfield  {author} {\bibinfo {author} {\bibfnamefont {Shi-Jie}\ \bibnamefont
  {Yang}}, \bibinfo {author} {\bibfnamefont {Quan-Sheng}\ \bibnamefont {Wu}},
  \bibinfo {author} {\bibfnamefont {Sheng-Nan}\ \bibnamefont {Zhang}}, \bibinfo
  {author} {\bibfnamefont {Shiping}\ \bibnamefont {Feng}}, \bibinfo {author}
  {\bibfnamefont {Wenan}\ \bibnamefont {Guo}}, \bibinfo {author} {\bibfnamefont
  {Yu-Chuan}\ \bibnamefont {Wen}}, \ and\ \bibinfo {author} {\bibfnamefont
  {Yue}\ \bibnamefont {Yu}},\ }\bibfield  {title} {\enquote {\bibinfo {title}
  {Generating ring dark solitons in an evolving {Bose-Einstein} condensate},}\
  }\href {\doibase 10.1103/PhysRevA.76.063606} {\bibfield  {journal} {\bibinfo
  {journal} {Phys. Rev. A}\ }\textbf {\bibinfo {volume} {76}},\ \bibinfo
  {pages} {063606} (\bibinfo {year} {2007})}\BibitemShut {NoStop}%
\bibitem [{\citenamefont {Gaunt}\ \emph {et~al.}(2013)\citenamefont {Gaunt},
  \citenamefont {Schmidutz}, \citenamefont {Gotlibovych}, \citenamefont
  {Smith},\ and\ \citenamefont {Hadzibabic}}]{PhysRevLett.110.200406}%
  \BibitemOpen
  \bibfield  {author} {\bibinfo {author} {\bibfnamefont {Alexander~L.}\
  \bibnamefont {Gaunt}}, \bibinfo {author} {\bibfnamefont {Tobias~F.}\
  \bibnamefont {Schmidutz}}, \bibinfo {author} {\bibfnamefont {Igor}\
  \bibnamefont {Gotlibovych}}, \bibinfo {author} {\bibfnamefont {Robert~P.}\
  \bibnamefont {Smith}}, \ and\ \bibinfo {author} {\bibfnamefont {Zoran}\
  \bibnamefont {Hadzibabic}},\ }\bibfield  {title} {\enquote {\bibinfo {title}
  {{Bose-Einstein Condensation of Atoms in a Uniform Potential}},}\ }\href
  {\doibase 10.1103/PhysRevLett.110.200406} {\bibfield  {journal} {\bibinfo
  {journal} {Phys. Rev. Lett.}\ }\textbf {\bibinfo {volume} {110}},\ \bibinfo
  {pages} {200406} (\bibinfo {year} {2013})}\BibitemShut {NoStop}%
\bibitem [{\citenamefont {Jaouadi}\ \emph {et~al.}(2010)\citenamefont
  {Jaouadi}, \citenamefont {Gaaloul}, \citenamefont {Viaris~de Lesegno},
  \citenamefont {Telmini}, \citenamefont {Pruvost},\ and\ \citenamefont
  {Charron}}]{PhysRevA.82.023613}%
  \BibitemOpen
  \bibfield  {author} {\bibinfo {author} {\bibfnamefont {A.}~\bibnamefont
  {Jaouadi}}, \bibinfo {author} {\bibfnamefont {N.}~\bibnamefont {Gaaloul}},
  \bibinfo {author} {\bibfnamefont {B.}~\bibnamefont {Viaris~de Lesegno}},
  \bibinfo {author} {\bibfnamefont {M.}~\bibnamefont {Telmini}}, \bibinfo
  {author} {\bibfnamefont {L.}~\bibnamefont {Pruvost}}, \ and\ \bibinfo
  {author} {\bibfnamefont {E.}~\bibnamefont {Charron}},\ }\bibfield  {title}
  {\enquote {\bibinfo {title} {{Bose-Einstein} condensation in dark power-law
  laser traps},}\ }\href {\doibase 10.1103/PhysRevA.82.023613} {\bibfield
  {journal} {\bibinfo  {journal} {Phys. Rev. A}\ }\textbf {\bibinfo {volume}
  {82}},\ \bibinfo {pages} {023613} (\bibinfo {year} {2010})}\BibitemShut
  {NoStop}%
\bibitem [{\citenamefont {Parker}\ \emph {et~al.}(2003)\citenamefont {Parker},
  \citenamefont {Proukakis}, \citenamefont {Leadbeater},\ and\ \citenamefont
  {Adams}}]{Parker_2003}%
  \BibitemOpen
  \bibfield  {author} {\bibinfo {author} {\bibfnamefont {N~G}\ \bibnamefont
  {Parker}}, \bibinfo {author} {\bibfnamefont {N~P}\ \bibnamefont {Proukakis}},
  \bibinfo {author} {\bibfnamefont {M}~\bibnamefont {Leadbeater}}, \ and\
  \bibinfo {author} {\bibfnamefont {C~S}\ \bibnamefont {Adams}},\ }\bibfield
  {title} {\enquote {\bibinfo {title} {Deformation of dark solitons in
  inhomogeneous {Bose{\textendash}Einstein} condensates},}\ }\href {\doibase
  10.1088/0953-4075/36/13/318} {\bibfield  {journal} {\bibinfo  {journal} {J.
  Phys. B, At. Mol. Opt. Phys.}\ }\textbf {\bibinfo {volume} {36}},\ \bibinfo
  {pages} {2891--2910} (\bibinfo {year} {2003})}\BibitemShut {NoStop}%
\bibitem [{\citenamefont {Cornish}\ \emph {et~al.}(2009)\citenamefont
  {Cornish}, \citenamefont {Parker}, \citenamefont {Martin}, \citenamefont
  {Judd}, \citenamefont {Scott}, \citenamefont {Fromhold},\ and\ \citenamefont
  {Adams}}]{Cornish2009}%
  \BibitemOpen
  \bibfield  {author} {\bibinfo {author} {\bibfnamefont {S.~L.}\ \bibnamefont
  {Cornish}}, \bibinfo {author} {\bibfnamefont {N.~G.}\ \bibnamefont {Parker}},
  \bibinfo {author} {\bibfnamefont {A.~M.}\ \bibnamefont {Martin}}, \bibinfo
  {author} {\bibfnamefont {T.~E.}\ \bibnamefont {Judd}}, \bibinfo {author}
  {\bibfnamefont {R.~G.}\ \bibnamefont {Scott}}, \bibinfo {author}
  {\bibfnamefont {T.~M.}\ \bibnamefont {Fromhold}}, \ and\ \bibinfo {author}
  {\bibfnamefont {C.~S.}\ \bibnamefont {Adams}},\ }\bibfield  {title} {\enquote
  {\bibinfo {title} {Quantum reflection of bright matter-wave solitons},}\
  }\href {\doibase 10.1016/j.physd.2008.07.011} {\bibfield  {journal} {\bibinfo
   {journal} {Physica D Nonlinear Phenomena}\ }\textbf {\bibinfo {volume}
  {238}},\ \bibinfo {pages} {1299--1305} (\bibinfo {year} {2009})}\BibitemShut
  {NoStop}%
\bibitem [{\citenamefont {Ruostekoski}\ \emph {et~al.}(2001)\citenamefont
  {Ruostekoski}, \citenamefont {Kneer}, \citenamefont {Schleich},\ and\
  \citenamefont {Rempe}}]{PhysRevA.63.043613}%
  \BibitemOpen
  \bibfield  {author} {\bibinfo {author} {\bibfnamefont {J.}~\bibnamefont
  {Ruostekoski}}, \bibinfo {author} {\bibfnamefont {B.}~\bibnamefont {Kneer}},
  \bibinfo {author} {\bibfnamefont {W.~P.}\ \bibnamefont {Schleich}}, \ and\
  \bibinfo {author} {\bibfnamefont {G.}~\bibnamefont {Rempe}},\ }\bibfield
  {title} {\enquote {\bibinfo {title} {Interference of a {Bose-Einstein}
  condensate in a hard-wall trap: From the nonlinear {Talbot} effect to the
  formation of vorticity},}\ }\href {\doibase 10.1103/PhysRevA.63.043613}
  {\bibfield  {journal} {\bibinfo  {journal} {Phys. Rev. A}\ }\textbf {\bibinfo
  {volume} {63}},\ \bibinfo {pages} {043613} (\bibinfo {year}
  {2001})}\BibitemShut {NoStop}%
\bibitem [{\citenamefont {Simula}\ \emph {et~al.}(2014)\citenamefont {Simula},
  \citenamefont {Davis},\ and\ \citenamefont
  {Helmerson}}]{PhysRevLett.113.165302}%
  \BibitemOpen
  \bibfield  {author} {\bibinfo {author} {\bibfnamefont {Tapio}\ \bibnamefont
  {Simula}}, \bibinfo {author} {\bibfnamefont {Matthew~J.}\ \bibnamefont
  {Davis}}, \ and\ \bibinfo {author} {\bibfnamefont {Kristian}\ \bibnamefont
  {Helmerson}},\ }\bibfield  {title} {\enquote {\bibinfo {title} {{Emergence of
  Order from Turbulence in an Isolated Planar Superfluid}},}\ }\href {\doibase
  10.1103/PhysRevLett.113.165302} {\bibfield  {journal} {\bibinfo  {journal}
  {Phys. Rev. Lett.}\ }\textbf {\bibinfo {volume} {113}},\ \bibinfo {pages}
  {165302} (\bibinfo {year} {2014})}\BibitemShut {NoStop}%
\bibitem [{\citenamefont {Groszek}\ \emph {et~al.}(2016)\citenamefont
  {Groszek}, \citenamefont {Simula}, \citenamefont {Paganin},\ and\
  \citenamefont {Helmerson}}]{PhysRevA.93.043614}%
  \BibitemOpen
  \bibfield  {author} {\bibinfo {author} {\bibfnamefont {Andrew~J.}\
  \bibnamefont {Groszek}}, \bibinfo {author} {\bibfnamefont {Tapio~P.}\
  \bibnamefont {Simula}}, \bibinfo {author} {\bibfnamefont {David~M.}\
  \bibnamefont {Paganin}}, \ and\ \bibinfo {author} {\bibfnamefont {Kristian}\
  \bibnamefont {Helmerson}},\ }\bibfield  {title} {\enquote {\bibinfo {title}
  {Onsager vortex formation in {Bose-Einstein} condensates in two-dimensional
  power-law traps},}\ }\href {\doibase 10.1103/PhysRevA.93.043614} {\bibfield
  {journal} {\bibinfo  {journal} {Phys. Rev. A}\ }\textbf {\bibinfo {volume}
  {93}},\ \bibinfo {pages} {043614} (\bibinfo {year} {2016})}\BibitemShut
  {NoStop}%
\bibitem [{\citenamefont {Huang}\ \emph {et~al.}(2003)\citenamefont {Huang},
  \citenamefont {Makarov},\ and\ \citenamefont {Velarde}}]{PRA.67.023604}%
  \BibitemOpen
  \bibfield  {author} {\bibinfo {author} {\bibfnamefont {Guoxiang}\
  \bibnamefont {Huang}}, \bibinfo {author} {\bibfnamefont {Valeri~A.}\
  \bibnamefont {Makarov}}, \ and\ \bibinfo {author} {\bibfnamefont {Manuel~G.}\
  \bibnamefont {Velarde}},\ }\bibfield  {title} {\enquote {\bibinfo {title}
  {Two-dimensional solitons in {B}ose-{E}instein condensates with a disk-shaped
  trap},}\ }\href {\doibase 10.1103/PhysRevA.67.023604} {\bibfield  {journal}
  {\bibinfo  {journal} {Phys. Rev. A}\ }\textbf {\bibinfo {volume} {67}},\
  \bibinfo {pages} {023604} (\bibinfo {year} {2003})}\BibitemShut {NoStop}%
\bibitem [{\citenamefont {Baizakov}\ \emph {et~al.}(2006)\citenamefont
  {Baizakov}, \citenamefont {Malomed},\ and\ \citenamefont
  {Salerno}}]{PRE.74.066615}%
  \BibitemOpen
  \bibfield  {author} {\bibinfo {author} {\bibfnamefont {Bakhtiyor~B.}\
  \bibnamefont {Baizakov}}, \bibinfo {author} {\bibfnamefont {Boris~A.}\
  \bibnamefont {Malomed}}, \ and\ \bibinfo {author} {\bibfnamefont {Mario}\
  \bibnamefont {Salerno}},\ }\bibfield  {title} {\enquote {\bibinfo {title}
  {Matter-wave solitons in radially periodic potentials},}\ }\href {\doibase
  10.1103/PhysRevE.74.066615} {\bibfield  {journal} {\bibinfo  {journal} {Phys.
  Rev. E}\ }\textbf {\bibinfo {volume} {74}},\ \bibinfo {pages} {066615}
  (\bibinfo {year} {2006})}\BibitemShut {NoStop}%
\bibitem [{\citenamefont {Cheng}\ \emph {et~al.}(2018)\citenamefont {Cheng},
  \citenamefont {Bai}, \citenamefont {Zhang}, \citenamefont {Xiong},\ and\
  \citenamefont {Yang}}]{LP.29.015501}%
  \BibitemOpen
  \bibfield  {author} {\bibinfo {author} {\bibfnamefont {Qiao-Ling}\
  \bibnamefont {Cheng}}, \bibinfo {author} {\bibfnamefont {Wen-Kai}\
  \bibnamefont {Bai}}, \bibinfo {author} {\bibfnamefont {Yao-Zhong}\
  \bibnamefont {Zhang}}, \bibinfo {author} {\bibfnamefont {Bo}~\bibnamefont
  {Xiong}}, \ and\ \bibinfo {author} {\bibfnamefont {Tao}\ \bibnamefont
  {Yang}},\ }\bibfield  {title} {\enquote {\bibinfo {title} {Influence of a
  dark soliton on the reflection of a {B}ose{\textendash}{E}instein condensate
  by a square barrier},}\ }\href {\doibase 10.1088/1555-6611/aaea78} {\bibfield
   {journal} {\bibinfo  {journal} {Laser Physics}\ }\textbf {\bibinfo {volume}
  {29}},\ \bibinfo {pages} {015501} (\bibinfo {year} {2018})}\BibitemShut
  {NoStop}%
\bibitem [{\citenamefont {Zhang}\ \emph {et~al.}(2011)\citenamefont {Zhang},
  \citenamefont {Hu}, \citenamefont {Wang}, \citenamefont {Liu},\ and\
  \citenamefont {Liu}}]{FOP.6.46}%
  \BibitemOpen
  \bibfield  {author} {\bibinfo {author} {\bibfnamefont {Xiao-fei}\
  \bibnamefont {Zhang}}, \bibinfo {author} {\bibfnamefont {Xing-hua}\
  \bibnamefont {Hu}}, \bibinfo {author} {\bibfnamefont {Deng-shan}\
  \bibnamefont {Wang}}, \bibinfo {author} {\bibfnamefont {Xun-xu}\ \bibnamefont
  {Liu}}, \ and\ \bibinfo {author} {\bibfnamefont {Wu-ming}\ \bibnamefont
  {Liu}},\ }\bibfield  {title} {\enquote {\bibinfo {title} {Dynamics of
  {B}ose-{E}instein condensates near {F}eshbach resonance in external
  potential},}\ }\href {\doibase 10.1007/s11467-010-0150-3} {\bibfield
  {journal} {\bibinfo  {journal} {Frontiers of Physics in China}\ }\textbf
  {\bibinfo {volume} {6}},\ \bibinfo {pages} {46 -- 60} (\bibinfo {year}
  {2011})}\BibitemShut {NoStop}%
\bibitem [{\citenamefont {Wang}\ \emph
  {et~al.}(2021{\natexlab{b}})\citenamefont {Wang}, \citenamefont {Xing},
  \citenamefont {Du}, \citenamefont {Xiong},\ and\ \citenamefont
  {Yang}}]{CPB.30.120303}%
  \BibitemOpen
  \bibfield  {author} {\bibinfo {author} {\bibfnamefont {Dong-Mei}\
  \bibnamefont {Wang}}, \bibinfo {author} {\bibfnamefont {Jian-Chong}\
  \bibnamefont {Xing}}, \bibinfo {author} {\bibfnamefont {Rong}\ \bibnamefont
  {Du}}, \bibinfo {author} {\bibfnamefont {Bo}~\bibnamefont {Xiong}}, \ and\
  \bibinfo {author} {\bibfnamefont {Tao}\ \bibnamefont {Yang}},\ }\bibfield
  {title} {\enquote {\bibinfo {title} {Quantum reflection of a
  {B}ose{\textendash}{E}instein condensate with a dark soliton from a step
  potential},}\ }\href {\doibase 10.1088/1674-1056/ac051e} {\bibfield
  {journal} {\bibinfo  {journal} {Chinese Physics B}\ }\textbf {\bibinfo
  {volume} {30}},\ \bibinfo {pages} {120303} (\bibinfo {year}
  {2021}{\natexlab{b}})}\BibitemShut {NoStop}%
\bibitem [{\citenamefont {Verma}\ \emph {et~al.}(2017)\citenamefont {Verma},
  \citenamefont {Rapol},\ and\ \citenamefont {Nath}}]{PRA.95.043618}%
  \BibitemOpen
  \bibfield  {author} {\bibinfo {author} {\bibfnamefont {Gunjan}\ \bibnamefont
  {Verma}}, \bibinfo {author} {\bibfnamefont {Umakant~D.}\ \bibnamefont
  {Rapol}}, \ and\ \bibinfo {author} {\bibfnamefont {Rejish}\ \bibnamefont
  {Nath}},\ }\bibfield  {title} {\enquote {\bibinfo {title} {Generation of dark
  solitons and their instability dynamics in two-dimensional condensates},}\
  }\href {\doibase 10.1103/PhysRevA.95.043618} {\bibfield  {journal} {\bibinfo
  {journal} {Phys. Rev. A}\ }\textbf {\bibinfo {volume} {95}},\ \bibinfo
  {pages} {043618} (\bibinfo {year} {2017})}\BibitemShut {NoStop}%
\bibitem [{\citenamefont {Haberichter}\ \emph {et~al.}(2019)\citenamefont
  {Haberichter}, \citenamefont {Kevrekidis}, \citenamefont
  {Carretero-Gonz\'alez},\ and\ \citenamefont {Edwards}}]{PRA.99.053619}%
  \BibitemOpen
  \bibfield  {author} {\bibinfo {author} {\bibfnamefont {M.}~\bibnamefont
  {Haberichter}}, \bibinfo {author} {\bibfnamefont {P.~G.}\ \bibnamefont
  {Kevrekidis}}, \bibinfo {author} {\bibfnamefont {R.}~\bibnamefont
  {Carretero-Gonz\'alez}}, \ and\ \bibinfo {author} {\bibfnamefont
  {M.}~\bibnamefont {Edwards}},\ }\bibfield  {title} {\enquote {\bibinfo
  {title} {Nonlinear waves in an experimentally motivated ring-shaped
  {B}ose-{E}instein-condensate setup},}\ }\href {\doibase
  10.1103/PhysRevA.99.053619} {\bibfield  {journal} {\bibinfo  {journal} {Phys.
  Rev. A}\ }\textbf {\bibinfo {volume} {99}},\ \bibinfo {pages} {053619}
  (\bibinfo {year} {2019})}\BibitemShut {NoStop}%
\bibitem [{\citenamefont {Yang}\ \emph {et~al.}(2014)\citenamefont {Yang},
  \citenamefont {Henning},\ and\ \citenamefont {Benedict}}]{JPB.47.035302}%
  \BibitemOpen
  \bibfield  {author} {\bibinfo {author} {\bibfnamefont {Tao}\ \bibnamefont
  {Yang}}, \bibinfo {author} {\bibfnamefont {Andrew~J}\ \bibnamefont
  {Henning}}, \ and\ \bibinfo {author} {\bibfnamefont {Keith~A}\ \bibnamefont
  {Benedict}},\ }\bibfield  {title} {\enquote {\bibinfo {title} {Bogoliubov
  excitation spectrum of an elongated condensate throughout a transition from
  quasi-one-dimensional to three-dimensional},}\ }\href {\doibase
  10.1088/0953-4075/47/3/035302} {\bibfield  {journal} {\bibinfo  {journal}
  {Journal of Physics B}\ }\textbf
  {\bibinfo {volume} {47}},\ \bibinfo {pages} {035302} (\bibinfo {year}
  {2014})}\BibitemShut {NoStop}%
\bibitem [{\citenamefont {Sciacca}\ \emph {et~al.}(2017)\citenamefont
  {Sciacca}, \citenamefont {Barenghi},\ and\ \citenamefont
  {Parker}}]{PhysRevA.95.013628}%
  \BibitemOpen
  \bibfield  {author} {\bibinfo {author} {\bibfnamefont {M.}~\bibnamefont
  {Sciacca}}, \bibinfo {author} {\bibfnamefont {C.~F.}\ \bibnamefont
  {Barenghi}}, \ and\ \bibinfo {author} {\bibfnamefont {N.~G.}\ \bibnamefont
  {Parker}},\ }\bibfield  {title} {\enquote {\bibinfo {title} {Matter-wave dark
  solitons in boxlike traps},}\ }\href {\doibase 10.1103/PhysRevA.95.013628}
  {\bibfield  {journal} {\bibinfo  {journal} {Phys. Rev. A}\ }\textbf {\bibinfo
  {volume} {95}},\ \bibinfo {pages} {013628} (\bibinfo {year}
  {2017})}\BibitemShut {NoStop}%
\bibitem [{\citenamefont {Clifford}\ \emph {et~al.}(1998)\citenamefont
  {Clifford}, \citenamefont {Arlt}, \citenamefont {Courtial},\ and\
  \citenamefont {Dholakia}}]{1998High}%
  \BibitemOpen
  \bibfield  {author} {\bibinfo {author} {\bibfnamefont {M.A}\ \bibnamefont
  {Clifford}}, \bibinfo {author} {\bibfnamefont {J}~\bibnamefont {Arlt}},
  \bibinfo {author} {\bibfnamefont {J}~\bibnamefont {Courtial}}, \ and\
  \bibinfo {author} {\bibfnamefont {K}~\bibnamefont {Dholakia}},\ }\bibfield
  {title} {\enquote {\bibinfo {title} {High-order {Laguerre-Gaussian} laser
  modes for studies of cold atoms},}\ }\href {\doibase
  https://doi.org/10.1016/S0030-4018(98)00464-7} {\bibfield  {journal}
  {\bibinfo  {journal} {Optics Communications}\ }\textbf {\bibinfo {volume}
  {156}},\ \bibinfo {pages} {300--306} (\bibinfo {year} {1998})}\BibitemShut
  {NoStop}%
\bibitem [{\citenamefont {Chin}\ \emph {et~al.}(2010)\citenamefont {Chin},
  \citenamefont {Grimm}, \citenamefont {Julienne},\ and\ \citenamefont
  {Tiesinga}}]{Rev.Mod.Phys.82.1225}%
  \BibitemOpen
  \bibfield  {author} {\bibinfo {author} {\bibfnamefont {Cheng}\ \bibnamefont
  {Chin}}, \bibinfo {author} {\bibfnamefont {Rudolf}\ \bibnamefont {Grimm}},
  \bibinfo {author} {\bibfnamefont {Paul}\ \bibnamefont {Julienne}}, \ and\
  \bibinfo {author} {\bibfnamefont {Eite}\ \bibnamefont {Tiesinga}},\
  }\bibfield  {title} {\enquote {\bibinfo {title} {Feshbach resonances in
  ultracold gases},}\ }\href {\doibase 10.1103/RevModPhys.82.1225} {\bibfield
  {journal} {\bibinfo  {journal} {Rev. Mod. Phys.}\ }\textbf {\bibinfo {volume}
  {82}},\ \bibinfo {pages} {1225--1286} (\bibinfo {year} {2010})}\BibitemShut
  {NoStop}%
\bibitem [{\citenamefont {Wang}\ \emph {et~al.}(2018)\citenamefont {Wang},
  \citenamefont {Dai}, \citenamefont {Wen}, \citenamefont {Liu}, \citenamefont
  {Jiang}, \citenamefont {Saito}, \citenamefont {Zhang},\ and\ \citenamefont
  {Zhang}}]{PRA.97.063607}%
  \BibitemOpen
  \bibfield  {author} {\bibinfo {author} {\bibfnamefont {Lin-Xue}\ \bibnamefont
  {Wang}}, \bibinfo {author} {\bibfnamefont {Chao-Qing}\ \bibnamefont {Dai}},
  \bibinfo {author} {\bibfnamefont {Lin}\ \bibnamefont {Wen}}, \bibinfo
  {author} {\bibfnamefont {Tao}\ \bibnamefont {Liu}}, \bibinfo {author}
  {\bibfnamefont {Hai-Feng}\ \bibnamefont {Jiang}}, \bibinfo {author}
  {\bibfnamefont {Hiroki}\ \bibnamefont {Saito}}, \bibinfo {author}
  {\bibfnamefont {Shou-Gang}\ \bibnamefont {Zhang}}, \ and\ \bibinfo {author}
  {\bibfnamefont {Xiao-Fei}\ \bibnamefont {Zhang}},\ }\bibfield  {title}
  {\enquote {\bibinfo {title} {Dynamics of vortices followed by the collapse of
  ring dark solitons in a two-component {B}ose-{E}instein condensate},}\ }\href
  {\doibase 10.1103/PhysRevA.97.063607} {\bibfield  {journal} {\bibinfo
  {journal} {Phys. Rev. A}\ }\textbf {\bibinfo {volume} {97}},\ \bibinfo
  {pages} {063607} (\bibinfo {year} {2018})}\BibitemShut {NoStop}%
\bibitem [{\citenamefont {Yang}\ \emph {et~al.}(2016)\citenamefont {Yang},
  \citenamefont {Hu}, \citenamefont {Zou},\ and\ \citenamefont
  {Liu}}]{SREP.6.29066}%
  \BibitemOpen
  \bibfield  {author} {\bibinfo {author} {\bibfnamefont {Tao}\ \bibnamefont
  {Yang}}, \bibinfo {author} {\bibfnamefont {Zhi-Qiang}\ \bibnamefont {Hu}},
  \bibinfo {author} {\bibfnamefont {Shan}\ \bibnamefont {Zou}}, \ and\ \bibinfo
  {author} {\bibfnamefont {Wu-Ming}\ \bibnamefont {Liu}},\ }\bibfield  {title}
  {\enquote {\bibinfo {title} {Dynamics of vortex quadrupoles in nonrotating
  trapped {Bose-Einstein} condensates},}\ }\href {\doibase 10.1038/srep29066}
  {\bibfield  {journal} {\bibinfo  {journal} {Scientific Reports}\ }\textbf
  {\bibinfo {volume} {6}},\ \bibinfo {pages} {29066} (\bibinfo {year}
  {2016})}\BibitemShut {NoStop}%
\end{thebibliography}
\end{document}